\documentclass[%
 nofootinbib,
 amsmath,amssymb,
 aps,
]{revtex4-2}

\usepackage{graphicx}% Include figure files
\usepackage{dcolumn}% Align table columns on decimal point
\usepackage{bm}% bold math
\usepackage{todonotes}
\usepackage{multirow}
\usepackage{booktabs}
\usepackage{adjustbox}
\usepackage{tabularx}
\usepackage{rotating}

\DeclareUnicodeCharacter{0301}{\'{e}}
\begin{document}

%\preprint{APS/123-QED}

\title{Pair distribution function analysis driven by atomistic simulations: Application to microwave radiation synthesized Ti$\text{O}_2$ and Zr$\text{O}_2$}% Force line breaks with \\

\author{Shuyan Zhang$^1$}
\author{Jie Gong$^1$}%
\author{Daniel Xiao$^2$}
\author{B. Reeja Jayan$^1$}
\author{Alan J. H. McGaughey$^{1}$}\email{mcgaughey@cmu.edu.}
\affiliation{%
 $^1$Department of Mechanical Engineering, Carnegie Mellon University, \\Pittsburgh, Pennsylvania 15213, USA  
}%
\affiliation{%
 $^2$Department of Materials Science and Engineering, Carnegie Mellon University, \\Pittsburgh, Pennsylvania 15213, USA  
}%

\begin{abstract}

A workflow is presented for performing pair distribution function (PDF) analysis of defected materials using structures generated from atomistic simulations. A large collection of structures, which differ in the types and concentrations of defects present, are obtained through energy minimization with an empirical interatomic potential. Each of the structures is refined against an experimental PDF. The structures with the lowest goodness of fit $R_w$ values are taken as being representative of the experimental structure. The workflow is applied to anatase titanium dioxide ($a$-TiO$_2$) and tetragonal zirconium dioxide ($t$-ZrO$_2$) synthesized in the presence of microwave radiation, a low temperature process that generates disorder. The results suggest that titanium vacancies and interstitials are the dominant defects in $a$-TiO$_2$, while oxygen vacancies dominate in $t$-ZrO$_2$. Analysis of the atomic displacement parameters extracted from the PDF refinement and mean squared displacements calculated from molecular dynamics simulations indicate that while these two quantities are closely related, it is challenging to make quantitative comparisons between them. The workflow can be applied to other materials systems, including nanoparticles.

\end{abstract}

\maketitle
%\tableofcontents
%\clearpage

\section{\label{sec:level1}Introduction}

The atomic structure of good quality crystalline materials can be obtained from X-ray crystallography, which only measures the Bragg peaks that result from the atomic periodicity. Crystallography alone is not sufficient for specifying the atomic structure of nanocrystalline and/or highly-defected samples ~\cite{christiansen2020there,yang2020structure, rietveld1969profile,egami2003underneath}. In pair distribution function (PDF) experiments, on the other hand, X-rays or neutrons are used and a wider angular range is detected so that both Bragg scattering and diffuse scattering are collected, the latter of which results from disorder, allowing for structural characterization without assuming periodicity~\cite{egami2003underneath,billinge2004beyond,billinge2019rise,nakamura2017unlocking}. The PDF is thus a powerful tool for quantitatively characterizing short-range and long-range atomic structure~\cite{nakamura2017unlocking}.

The PDF, denoted by $G(r)$,  provides the scaled probability of finding two atoms a distance of $r$ apart~\cite{yang2020structure}. An experimentally-measured PDF can be analyzed by adjusting the parameters of an assumed structure model, such as the lattice constants, atomic positions, and grain/particle size. The structure is refined in real-space by minimizing the difference between its PDF and the experimental PDF~\cite{christiansen2020there, yang2020structure}. The refinement process to perform this analysis has been implemented in PDFgui~\cite{farrow2007pdffit2}, DiffPy-cmi~\cite{juhas2015complex}, and TOPAS~\cite{coelho2015fast}. 

A major challenge in PDF modeling is the selection of the starting atomic structure~\cite{christiansen2020there,yang2020structure,chapman2015applications}. Significant information about the sample (e.g., the crystal phase) is required to achieve satisfactory results. Typically, PDF modeling involves manual trial-and-error refinement of multiple structure models~\cite{chapman2015applications,banerjee2020cluster}. There have been attempts to automate and accelerate this process, where a large number of structures are either pulled from a materials database or generated automatically~\cite{yang2020structure,banerjee2020cluster,boullay2014fast,lutterotti2019full}. Such approaches can be efficient for the identification of the crystal structure of an unknown material. They may not be sufficient, however, when the material and phase are already known and detailed structural information is required. For example, if the sample is disordered and/or a nanoparticle, where there can be atomic displacements away from the perfect bulk structure due to defects and/or surfaces~\cite{mishra2021point}. Since the crystal structures available in most databases are for perfect bulk phases~\cite{downs2003american,belsky2002new,jain2013commentary}, a method for generating candidate structures that takes into account the atomic displacements induced by disorder and/or surfaces is needed.

Atomistic simulations provide a solution to this challenge. While density functional theory (DFT) can perform first principles-based  energy calculations, it is limited to small systems with minimal complexity by its large computational cost. Empirical interatomic potentials, on the other hand, can efficiently provide the total energy of a large, complex system as an explicit function of its atomic coordinates~\cite{torrens2012interatomic}. A well-parameterized potential can maintain high accuracy when compared to a DFT calculation.

In minimizing the energy of a system, the atomic positions will change to account for perturbations to the perfect structure (e.g., defects). Structure models generated via energy minimization with an interatomic potential can therefore be used to represent complex systems that are not available in materials databases. Furthermore, an energy minimization-based approach is advantageous compared to the reverse Monte Carlo (RMC) method. In RMC, atoms are allowed to move under certain constraints (e.g., bond angles, bond lengths) until the difference between the calculated and experimental structural characteristic of interest (e.g, the PDF) is minimized~\cite{mcgreevy2001reverse}. Incorrect structures, however, can be generated if the appropriate combinations of such constraints are not applied~\cite{opletal2017reverse}.

Defect-induced changes in the local atomic environment can be described by atomistic calculations if the crystal phase and defect information (e.g., types, concentrations) are specified. Herein, we describe a PDF analysis workflow for modeling defected crystals. Structures with different types and concentrations of defects are first created and then relaxed using energy minimization. The PDF of each relaxed structure is then refined against the experimental PDF. The defect types and concentrations are then estimated from the structures that yield refined PDFs with the smallest differences compared to the experimental PDF.

Measured PDFs of anatase titanium dioxide (TiO$_2$) and tetragonal zirconium dioxide (ZrO$_2$) grown using low temperature solution synthesis in the presence of microwave radiation (MWR) are used to demonstrate this method. External fields can induce defects that alter the diffusion of space charges across grain boundaries and ultimately lead to efficient processing~\cite{raj2011influence, jha2019defect, jongmanns2020element,nakamura2021linking}. Our goal is to identify the types and concentrations of point defects that are present in MWR-grown anatase TiO$_2$ and tetragonal ZrO$_2$. While we focus on structures with local disorder, this workflow can also benefit PDF analysis of nanoparticles~\cite{petkov2002structure} and perovskites~\cite{krayzman2008simultaneous}, which are challenging to characterize using conventional methods.

The rest of the paper is organized as follows. In Secs.~\ref{structure model generation} and~\ref{pdf analysis}, the atomistic calculations, structure generation, and PDF analysis are described. In Sec.~\ref{potential selection}, the interatomic potential selection process is presented, which is guided by predictions of lattice constants and elastic constants. In Sec.~\ref{DFE-results}, defect formation energy predictions are compared with literature values. The PDF analysis is presented in Sec.~\ref{S-pdf}. We find that Ti vacancy and Ti interstitial are the dominant defects in MWR-grown anatase TiO$_2$ and that O vacancy is the dominant defect in MWR-grown tetragonal ZrO$_2$.

\section{Methods}

\subsection{\label{structure model generation}Atomistic calculations and structure model generation}

The nature of the bonding in Ti$\text{O}_2$ and Zr$\text{O}_2$, which have relatively strong covalent characteristics, makes it challenging to model the atomic interactions using an interatomic potential. We also require that the potential be able to handle heterogeneous, defected systems. Atomistic calculations of materials with electrostatic interactions typically maintain fixed charges on the ions. Charge transfer may happen around a defect site, however, causing deviations of the electrostatic energies and particle positions~\cite{ufimtsev2011charge,campbell1999dynamics}. A variable-charge potential allows for the redistribution of charges based on the local environment~\cite{ogata1999variable,shan2011molecular,ogata2000role}. We identified two candidate variable charge potentials: second moment tight-binding charge equilibrium (SMTB-Q)~\cite{tetot2008tight,maras2015improved} and reactive force field (ReaxFF)~\cite{van2008reaxff,kim2013development,huygh2014development}, which are evaluated in Sec.~\ref{potential selection}.

All atomistic calculations are performed using the Large-scale Atomic/Molecular Massively Parallel Simulator (LAMMPS)~\cite{plimpton1995fast}. The space groups, supercell sizes, and number of atoms in the simulation boxes of anatase and rutile Ti$\text{O}_2$, and tetragonal, cubic, and monoclinic Zr$\text{O}_2$ are listed in Table~\ref{tab:setup}. While we are focused on anatase Ti$\text{O}_2$ and tetragonal Zr$\text{O}_2$, the other phases were included to best assess the potentials. Periodic boundary conditions are applied in all three directions. The long-range electrostatic interaction is calculated using the Wolf summation method~\cite{tetot2008tight, wolf1999exact}. The time step for the molecular dynamics (MD) simulations is 0.2 fs. The charge on each atom is updated every time step using the charge equilibrium (Qeq) scheme developed by Rappé and Goddard ~\cite{rappe1991charge}. 

We built structures with different types and concentrations of randomly-placed point defects and relaxed them at zero temperature by energy minimization. To investigate the sensitivity of the refined PDF to the defect locations, ten structures were created for each configuration. We consider common point defects for both the cation (Ti and Zr) and the anion (O): vacancies, interstitials, Frenkel pairs (a cation vacancy and interstitial pair), and anti-Frenkel pairs (an anion vacancy and interstitial pair), and combinations of these defect types.

\begin{table}[t]
\caption{\label{tab:setup}Space groups, supercell sizes, and number of atoms in the Ti$\text{O}_2$ and Zr$\text{O}_2$ simulation boxes. $a$, $b$, and $c$ are the lattice constants.}
\begin{ruledtabular}
\begin{tabular}{lccc}
 \textbf{Phase} & \textbf{Space Group} & \textbf{Supercell} & \textbf{Number of Atoms} \\ \hline
Anatase Ti$\text{O}_2$    & $I41/amd$              & $7a \times 7a \times 3c$       & 1764                     \\
Rutile Ti$\text{O}_2$     & $P42/mnm$              & $6a \times 6a \times 9c$       & 1944                     \\
Tetragonal Zr$\text{O}_2$ & $P42/nmc$              & $7a \times 7a \times 5c$       & 1470                     \\
Cubic Zr$\text{O}_2$      & $Fm3m$                 & $5a \times 5a \times 5a$        & 1500                     \\ 
Monoclinic Zr$\text{O}_2$ & $P21/c$                & $5a \times 5b \times 5c$       & 1500                     \\
\end{tabular}
\end{ruledtabular}
\end{table}

\subsection{\label{pdf analysis}Pair distribution function (PDF) analysis}

We use measured PDFs for MWR-grown anatase TiO$_2$ and tetragonal ZrO$_2$ thin films from previous studies~\cite{nakamura2017unlocking,nakamura2021linking}. The data acquisition was performed at the X-ray Powder Diffraction Beamline, 28-ID-2, at the National Synchrotron Light Source II at Brookhaven National Laboratory. 

The PDF refinement is performed using the DiffPy-cmi package~\cite{juhas2015complex}. The structure models are fit against the measured PDFs by refining global and phase-specific scaling factors, lattice parameters, isotropic atomic displacement parameters (ADPs, denoted by $U_{iso}$), a low-$r$ peak sharpening coefficient for correlated motion of nearby atoms, and a PDF peak envelope function, which dampens the signal as a function of separation to account for grain/particle size. The quality of the fit is quantified by a goodness-of-fit value $R_w$, which is  calculated as~\cite{egami2003underneath}

\begin{equation}
 R_{\mathrm{w}}=\sqrt{\frac{\sum_{n}\left(G_{\mathrm{obs}, n}-G_{\mathrm{calc}, n}\right)^{2}}{\sum_{n} G_{\mathrm{obs}, n}^{2}},}
\end{equation}
where $G_{\mathrm{obs}, n}$ is the $n$th point on the experimentally measured PDF and $G_{\mathrm{calc}, n}$ is the $n$th point on the refined PDF. The PDF refinement is done from 1.5 \AA~to 30 \AA~in an increment of 0.01 \AA. Periodic boundary conditions are applied to the structure models~\cite{xyzfiles}. This step is important because if the supercell is not periodic, the PDF peaks will dampen as the atomic separation approaches the simulation box size.\par

\section{Potential Selection and Assessment}

\subsection{\label{potential selection}Lattice constants and elastic constants}
We performed atomistic calculations using  SMTB-Q and ReaxFF for both TiO$_2$ and ZrO$_2$. The SMTB-Q potential~\cite{maras2015improved} and two ReaxFF parametrizations, i.e., ReaxFF – Ti$\text{O}_2$/$\text{H}_2$O~\cite{kim2013development} and ReaxFF – defect~\cite{huygh2014development} for Ti$\text{O}_2$ all produce stable structures in MD simulations and will be further considered in the lattice constant and elastic constant calculations. The ReaxFF parametrized for yttria-stabilized zirconia~\cite{van2008reaxff}, however, does not generate stable structures in MD simualtions for the tetragonal, cubic, and monoclinic phases of Zr$\text{O}_2$. The SMTB-Q potential for Zr$\text{O}_2$~\cite{tetot2008tight} does generate stable structures for perfect and defected phases and will be used in the subsequent analysis.

The lattice constants and elastic constants of anatase Ti$\text{O}_2$ calculated using the SMTB-Q and ReaxFF are presented in Table~\ref{tab:elastic} along with experimental values from the literature~\cite{rahimi2016review,yao2007ab}. The calculation details are provided in Appendix~\ref{app:prop pred}. The lattice parameters predicted by all three potentials are close to the experimental values with a largest deviation of 1.7$\%$. The accuracy of the elastic constants is quantified by the root-mean-square error (RMSE) of the unique elements of the elastic constant tensor compared to the experimental values. The elastic constant RMSE from the SMTB-Q potential is 17.9 GPa, thirteen and five times smaller than those predicted by ReaxFF – Ti$\text{O}_2$/$\text{H}_2$O and ReaxFF – defect. In addition, the elastic constants calculated by SMTB-Q maintain the crystal symmetry (i.e., $C_{11}$ = $C_{22}$, $C_{13}$ = $C_{23}$, and $C_{44}$ = $C_{55}$), which is not the case with those calculated by ReaxFF – Ti$\text{O}_2$/$\text{H}_2$O and ReaxFF – defect. Based on the lattice constant and elastic constant results, we decided to use the SMTB-Q potential in our further calculations of Ti$\text{O}_2$. The lattice constants and elastic constants of rutile Ti$\text{O}_2$ are provided in Table S1 of the Supplemental Material~\cite{SM}. 

The lattice constants and elastic constants of tetragonal Zr$\text{O}_2$ calculated using the SMTB-Q potential are also reported in Table~\ref{tab:elastic}.  Although the ZrO$_2$ elastic constants have a larger RMSE compared to that for Ti$\text{O}_2$, the lattice constants are well predicted with an average error of 2.4\% compared to the experimental values~\cite{igawa1993crystal,kisi1998elastic}. The elastic constants that represent the reaction to a normal stress (i.e., $C_{11}$, $C_{22}$, and $C_{33}$) are reasonably accurate, with an average error of 18\%. The lattice constants and elastic constants of monoclinic and cubic Zr$\text{O}_2$ are provided in Table S2 of the Supplemental Material~\cite{SM}.

\begin{table*}
\caption{\label{tab:elastic}Lattice constants ($a$, $c$) (\AA), elastic constants ($c_{ij}$) (GPa), bulk modulus ($B_\text{V}$) (GPa), and shear modulus ($G_\text{V}$) (GPa) calculated using the SMTB-Q and ReaxFF potentials, and literature experimental values.}
\begin{ruledtabular}
\begin{tabular}{lcccc|cc}
 & \multicolumn{4}{c}{\bf Anatase Ti$\text{O}_2$}&\multicolumn{2}{c}{\bf Tetragonal Zr$\text{O}_2$}\\\cline{2-7}
 &SMTB-Q &ReaxFF-Ti$\text{O}_2$/$\text{H}_2\text{O}$& ReaxFF-defect &Exp.~\cite{rahimi2016review,yao2007ab} & SMTB-Q &Exp.~\cite{igawa1993crystal,kisi1998elastic} \\ \hline
 $a$ &3.809 &3.798 &3.832 &3.784 &3.681 &3.591\\
 $c$ &9.597 &9.550 &9.673 &9.515 &5.285 &5.169\\  \hline
 $c_{11}$ &341 &401 &402 &337 &294 &327\\
 $c_{22}$ &341 &401 &412 &337 &294 &327\\
 $c_{33}$ &198 &211 &245 &192 &178 &264\\
 $c_{12}$ &104 &87 &101 &139 &191 &100\\
 $c_{13}$ &103 &115 &145 &136 &16.5 &62\\
 $c_{23}$ &103 &113 &155 &136 &16.5 &62\\
 $c_{44}$ &56 &-675 &-238 &54 &41.7 &59\\
 $c_{55}$ &56 &-58 &90 &54 &41.7 &59\\
 $c_{66}$ &63 &35 &53 &60 &181 &64\\
 $c_{ij}$ RMSE &17.9 &249 &106 &- &190 &-\\\hline
 $B_\text{V}$\footnote{Eq.~\eqref{E-BV}} &167 &182 &207 &178 &135 &152\\
 $G_\text{V}$\footnote{Eq.~\eqref{E-GV}} &58 &-32 &-32 &57 & 89 & 83\\
\end{tabular}
\end{ruledtabular}
\end{table*}

\subsection{\label{DFE-results}Defect formation energy}

\begin{figure*}[h]
    \centering
    \includegraphics[width=0.93\textwidth]{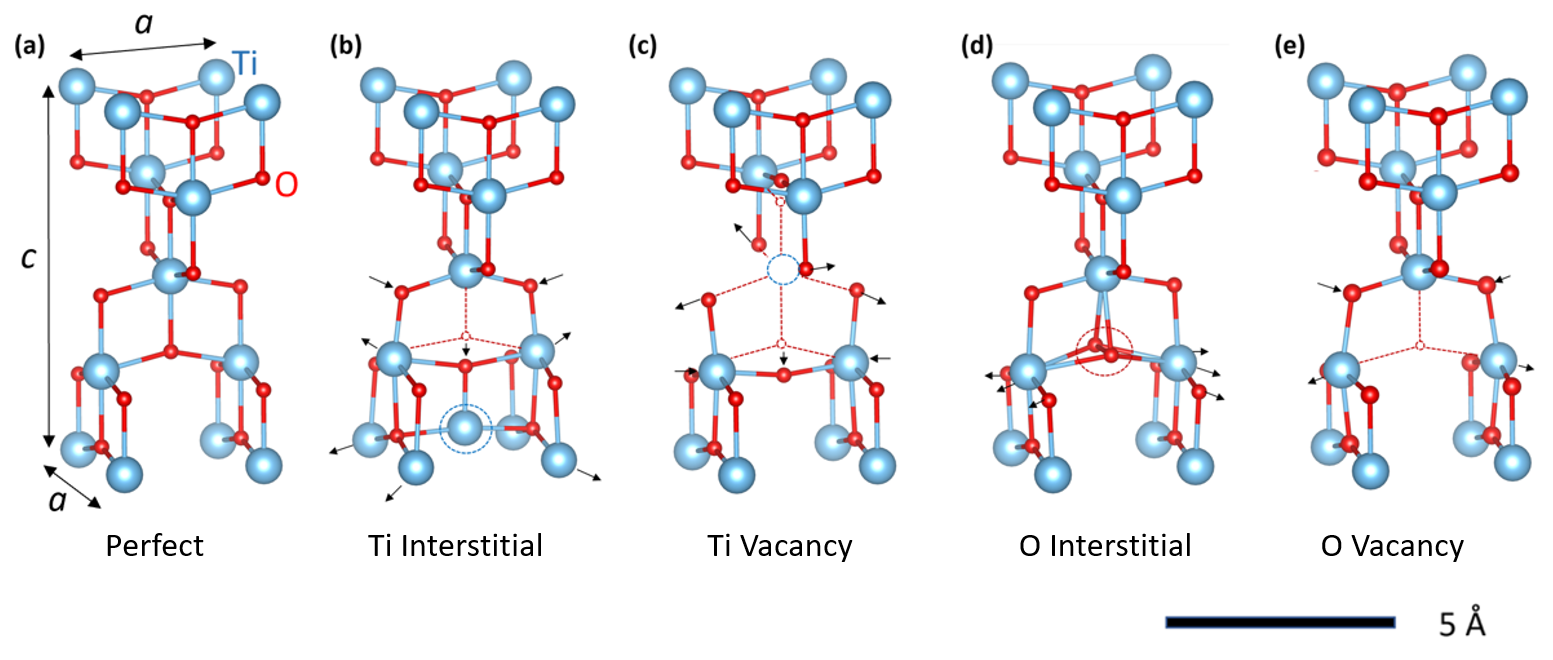}
    \caption{\label{fig_structure_TiO2} Relaxed atomic structures of anatase TiO$_2$ from the SMTB-Q potential: (a) perfect, (b) Ti interstitial, (c) Ti vacancy, (d) O interstitial, and (e) O vacancy. The dashed circles indicate the defect sites. The arrows indicate the direction of displacements after relaxation. The structures are qualitatively similar to DFT-predicted structures~\cite{na2006first}. }
\end{figure*}

The relaxed point defect structures predicted by SMTB-Q in anatase Ti$\text{O}_2$ and tetragonal ZrO$_2$, the phases that are grown under MWR, are shown in Figs.~\ref{fig_structure_TiO2}(a)$\textendash$1(e) and \ref{fig_structure_ZrO2}(a)$\textendash$2(e). The calculation details are presented in Appendix~\ref{app:prop pred}. The resulting atomic displacements around the point defects in Ti$\text{O}_2$ (denoted by arrows) are similar to those observed in previous DFT calculations~\cite{na2006first} and range from 0.1~\AA~ (O interstitial) to 0.4~\AA~ (Ti vacancy).  For Zr$\text{O}_2$, the atomic displacements around a point defect are less than 0.0001~\AA~ and are not shown. 

\begin{figure*}[h]
    \centering
    \includegraphics[width=0.95\textwidth]{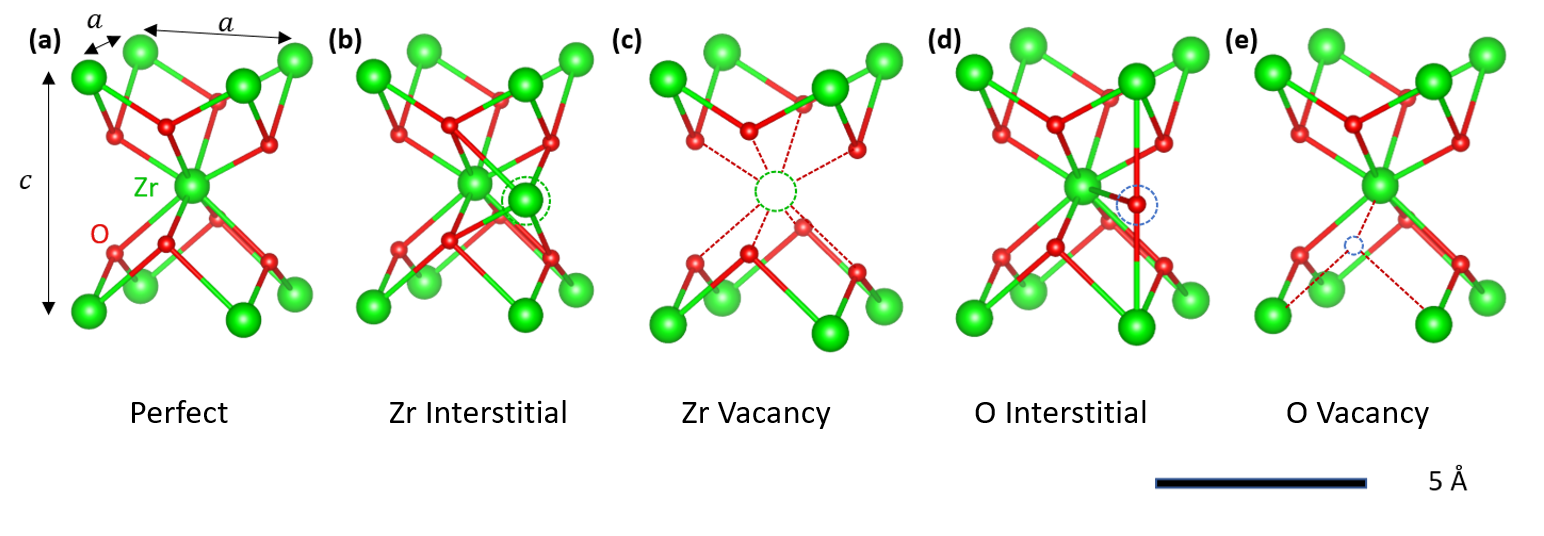}
    \caption{\label{fig_structure_ZrO2} 
    Relaxed atomic structures of tetragonal ZrO$_2$ from the SMTB-Q potential: (a) perfect, (b) Zr interstitial, (c) Zr vacancy, (d) O interstitial, and (e) O vacancy. The dashed circles indicate the defect sites. The atomic displacements around a point defect are less than 0.0001~\AA~ and are not shown.}
\end{figure*}

The anatase Ti$\text{O}_2$ defect formation energies predicted by SMTB-Q and from previous DFT calculations are provided in Table~\ref{tab:TiO2&ZrO2}. Also provided are the tetragonal Zr$\text{O}_2$ defect formation energies predicted by SMTB-Q, previous DFT calculations, and previous calculations for the polarizable Born-Mayer (BM) core-shell potential~\cite{dwivedi1990computer}. In the DFT studies, the defect formation energies depend on the charge states, the elemental chemical potentials, the Fermi level, and the valence-band maximum of the perfect crystal. As such, several discrete values are often reported for a given defect type. The full range is provided in Table~\ref{tab:TiO2&ZrO2}. The defect formation energies for other phases of TiO$_2$ and ZrO$_2$ are provided in Tables S3 and S4 of the Supplemental Material~\cite{SM}.

For anatase TiO$_2$, the SMTB-Q defect formation energies fall within the DFT energy ranges except for the O interstitial (0.07 eV above the maximum DFT value). SMTB-Q predicts that the smallest formation energy is for a Zr interstitial. For tetragonal Zr$\text{O}_2$, the SMTB-Q defect formation energies fall close to the DFT results except for the O interstitial. All SMTB-Q results are in better agreement with the DFT results than the BM potential. SMTB-Q predicts that the smallest formation energy is for a Ti interstitial.

\begin{table*}
\caption{\label{tab:TiO2&ZrO2}Defect formation energies in eV for anatase Ti$\text{O}_2$ and tetragonal Zr$\text{O}_2$ from the SMTB-Q potential, and previous DFT and BM potential calculations.}
\begin{ruledtabular}
\begin{tabular}{lccccc}
&\multicolumn{2}{c}{\bf Anatase Ti$\text{O}_2$}&\multicolumn{3}{c}{\bf Tetragonal Zr$\text{O}_2$}\\\cline{2-6}
 &SMTB-Q &DFT &SMTB-Q &DFT~\cite{youssef2012intrinsic} &BM~\cite{dwivedi1990computer}\\ \hline
 $\text{O Vacancy}$ &1.85 &0.70$\textendash$4.23~\cite{morgan2010intrinsic} & 4.42   & -0.76 $\textendash$ 6.10 & 15.62\\
 $\text{Cation Vacancy}$ &2.93 &-0.7 $\textendash$ 5.80~\cite{na2006first} & 9.63   & 6.11 & 85.04\\
 $\text{O Interstitial}$ &4.37 &0.60 $\textendash$ 4.30~\cite{na2006first} & 7.74   & 1.79 & -10.42\\
 $\text{Cation Interstitial}$ &1.07 &0.70 $\textendash$ 7.74~\cite{morgan2010intrinsic} & 1.85   & 1.96 & -67.41\\

 $\text{O Frenkel Pair}$ &3.86 &-- & 11.39  & 8.07 $\textendash$ 18.57    & 17.63\\
 $\text{Cation Frenkel Pair}$ &5.38 &-- & 12.12  & 4.11 $\textendash$ 7.41    & 5.20\\
\end{tabular}
\end{ruledtabular}
\end{table*}

%\clearpage

\section{Pair distribution function (PDF) analysis\label{S-pdf}}

\subsection{\label{pdf-tio2}Anatase Ti$\text{O}_2$}
\begin{figure}[b]
    \centering
    \includegraphics[scale=0.35]{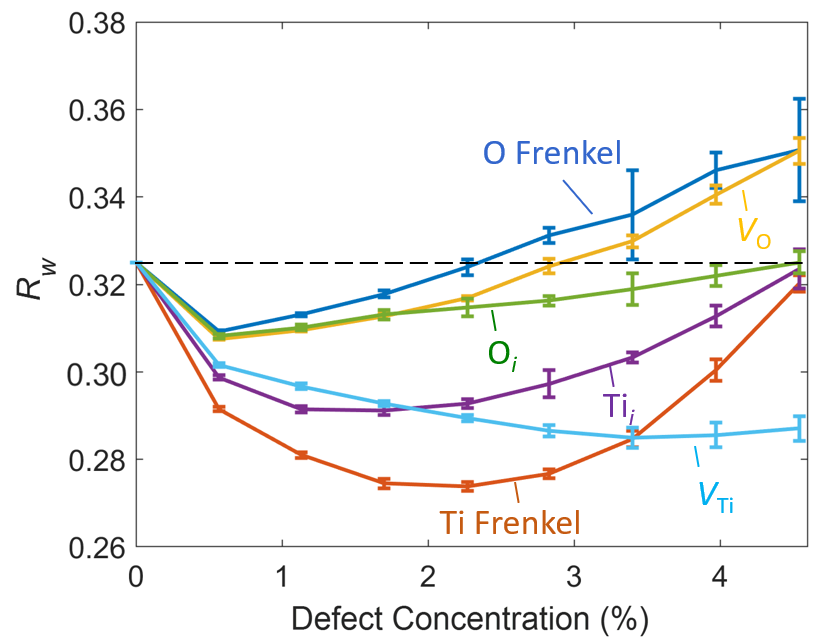}
    \caption{\label{fig:rw_vs_c}Variation of $R_w$ with defect concentration for anatase Ti$\text{O}_2$. The error bars represent the standard deviation from ten randomly-generated structures for each defect type and concentration.}
\end{figure}

We now perform PDF analysis with starting structures obtained as described in Sec.~\ref{structure model generation} to determine the defect types (vacancies, interstitials, Frenkel pairs, anti-Frenkel pairs) and concentrations present in MWR-grown anatase Ti$\text{O}_2$ (this section) and tetragonal Zr$\text{O}_2$ (Sec.~\ref{pdf-zro2}). The variation of $R_w$ with defect concentration for different defects in anatase Ti$\text{O}_2$ is plotted in Fig.~\ref{fig:rw_vs_c}. The concentration of a defect type is specified by dividing the number of that type of defect by the total number of atoms before defects are inserted. The mean values and standard deviations (as shown by the error bars) are obtained from ten randomly generated structures for each defect type and concentration. 

For all six defect types, $R_w$ first decreases starting from 0.325 (the perfect structure, plotted as a dashed horizontal line), reaches a minimum, and then increases as the defect concentration varies from 0 to 4.5\%. Introducing Ti defects improves the PDF refinement more than O defects as indicated by the lower $R_w$ across all concentrations. The small error bars indicate that $R_w$ is insensitive to the defect configuration except at the higher concentrations, when clustering is possible. The lowest $R_w$ (0.274) comes from the structure with 2.3\% Ti Frenkel pairs. 

\begin{figure}[h]
    \centering
    \includegraphics[scale=0.5]{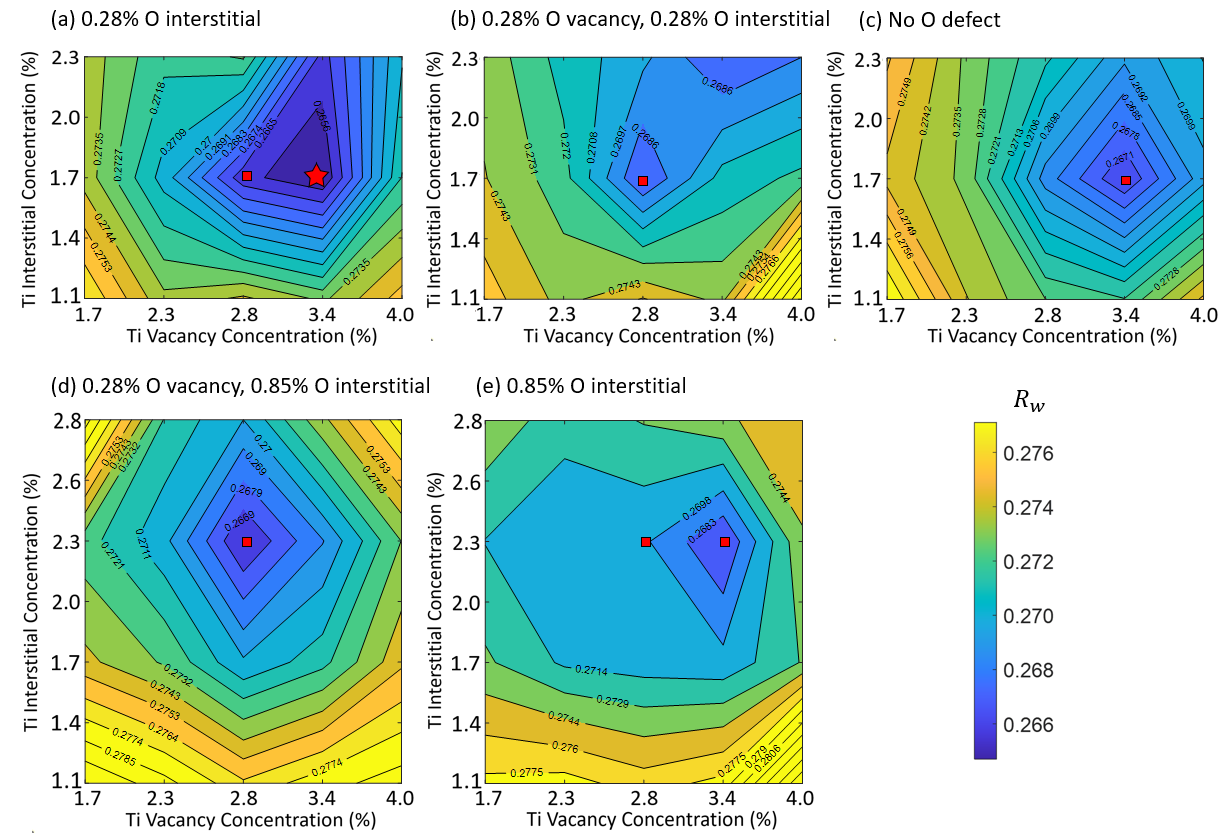}
    \caption{\label{fig:contour}Variation of anatase TiO$_2$ $R_w$ with Ti and O defect concentrations. The red markers indicate the seven lowest $R_w$ values. The lowest $R_w$, the red star in (a), results from the PDF refinement of a structure model with  3.40\% Ti vacancies, 1.70\% Ti interstitials and 0.28\% O interstitials.}
\end{figure}

\begin{figure*}[h]
    \centering
    \includegraphics[scale=0.46]{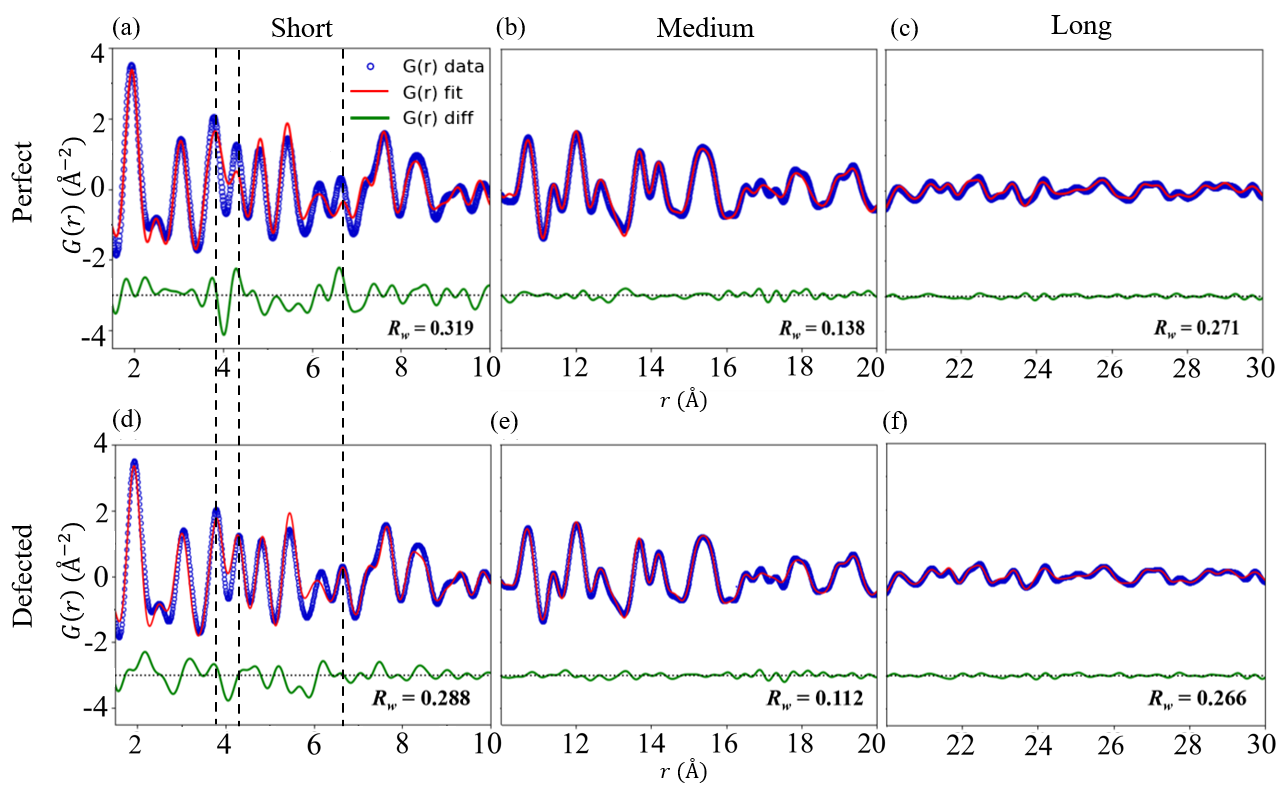}
    \caption{\label{fig:splitted}PDF refinement for anatase Ti$\text{O}_2$ separated into short, medium, and long range order. The green curves are the differences between the refined and experimental PDFs. (a)-(c) are the refined PDFs using the perfect structure and (d)-(f) are the refined PDFs using the the defected structure with the lowest $R_w$.}
\end{figure*}

\begin{table}[h!]
\caption{\label{tab:240 structures}Defect concentrations sampled for anatase Ti$\text{O}_2$ structures.}
\begin{ruledtabular}
\begin{tabular}{lc}
  \textbf{Defect Type} & \textbf{Concentrations (\%)}\\ \hline
Ti Vacancy ($V_\text{Ti}$)         & 1.70, 2.27, 2.83, 3.40, 4.00 \\
Ti Interstitial ($\text{Ti}_i$)    & 1.13, 1.70, 2.28, 2.83\\
O Vacancy ($V_\text{O}$)           & 0, 0.28, 0.57, 0.85\\
O Interstitial ($\text{O}_i$)     &  0, 0.28, 0.57, 0.85
\end{tabular}
\end{ruledtabular}
\end{table}

Based on this result, we conducted a more comprehensive study. The concentrations of the four types of point defect (Ti vacancy, Ti interstitial, O vacancy, and O interstitial) were simultaneously varied and the PDFs of the resulting structure models were refined against the experimental PDF. We considered 240 defect type combinations and generated ten random structures for each combination. The concentrations for each defect type are provided in Table~\ref{tab:240 structures}. We then plotted the average $R_w$ values as two-dimensional contours based on the Ti vacancies and Ti interstitials, for a constant concentration of O defects. We considered sixteen combinations of O defects (Table~\ref{tab:240 structures}), resulting in sixteen contours. The five contour plots containing the lowest $R_w$ values are shown in Figs.~\ref{fig:contour}(a)-\ref{fig:contour}(e) and the others are presented in Figs. S1 and S2 of the Supplemental Material~\cite{SM}. The lowest $R_w$ (0.264) corresponds to the structure with 3.40\% Ti vacancies, 1.70\% Ti interstitials, and 0.28\% O interstitials [indicated by the red star marker in Fig.~\ref{fig:contour}(a)]. The refinement parameters associated with this structure are listed in Table S5 of the Supplemental Material~\cite{SM}. Six other defected structures give a similar $R_w$ (within 3\% of 0.264) after the PDF refinement [indicated by red square markers in Figs.~\ref{fig:contour}(a)-\ref{fig:contour}(e)]. All of these structures have the Ti vacancy as the dominant defect type, followed by the Ti interstitial.

The PDFs of the perfect anatase phase and the defected anatase structure with the lowest $R_w$ (both refined against the same experimental PDF) are provided in Figs.~\ref{fig:splitted}(a)\textendash\ref{fig:splitted}(f), where they are split into short range (0 $\textendash$ 10 \AA), medium range (10 $\textendash$ 20 \AA),  and long range (20 $\textendash$ 30 \AA) order. The short range $R_w$ decreases from 0.319 to 0.288 and the medium range $R_w$ decreases from 0.138 to 0.112. There are notable improvements for the peaks at 4.3 \AA and 6.7 \AA, both of which represent Ti$\textendash$O pairs as identified by the dashed vertical lines. Improvement can also be seen for the Ti$\textendash$Ti peak at 3.8 \AA. The corresponding atom pairs are shown in Fig. S3(a) of the Supplemental Material~\cite{SM}. The improved fit to Ti$\textendash$O and Ti$\textendash$Ti peaks suggest that the defected model, which is dominated by Ti vacancies and interstitials, is a good representation of the defected structure. The refinement results for the other six structures with low $R_w$ values have improved fitting at the same distances.  The refinement to the long range leads to a smaller $R_w$ decrease from 0.271 to 0.266. This result suggests that despite the local disorder, the atomic structure of the MWR-grown Ti$\text{O}_2$ remains anatase, which agrees with previous work by Nakamura et al$.$ on the same experimental sample~\cite{nakamura2017unlocking}. The introduction of the defects makes the local atomic environment in the lower $r$ region similar to that in the MWR-grown Ti$\text{O}_2$. 

The SMTB-Q results provided in Table~\ref{tab:TiO2&ZrO2} indicate that Ti interstitial has the lowest formation energy, followed by O vacancy, Ti vacancy, and O interstitial. The low-ends of the DFT ranges indicate that Ti vacancy has the lowest formation energy, followed by approximately the same value for O interstitial, Ti interstitial, and O vacancy. The defect concentrations predicted for the MWR-grown Ti$\text{O}_2$ follow a different order, with a dominance of Ti vacancy and Ti interstitial. This finding may be due to the fact that the defect formation energies are calculated at zero temperature without considering the electromagnetic field, while the experimental sample is synthesized under MWR at 140 $\textendash$ 160 $^\circ$C. %As the potential reasons discussed in Sec.~\ref{DFE-results} for the difference between defect formation energies from DFT and the SMTB-Q potential, the charge state of the defect and the experimental conditions can also affect the results. 

\subsection{\label{pdf-zro2}Tetragonal Zr$\text{O}_2$}
\begin{figure}[h]
    \centering
    \includegraphics[scale=0.31]{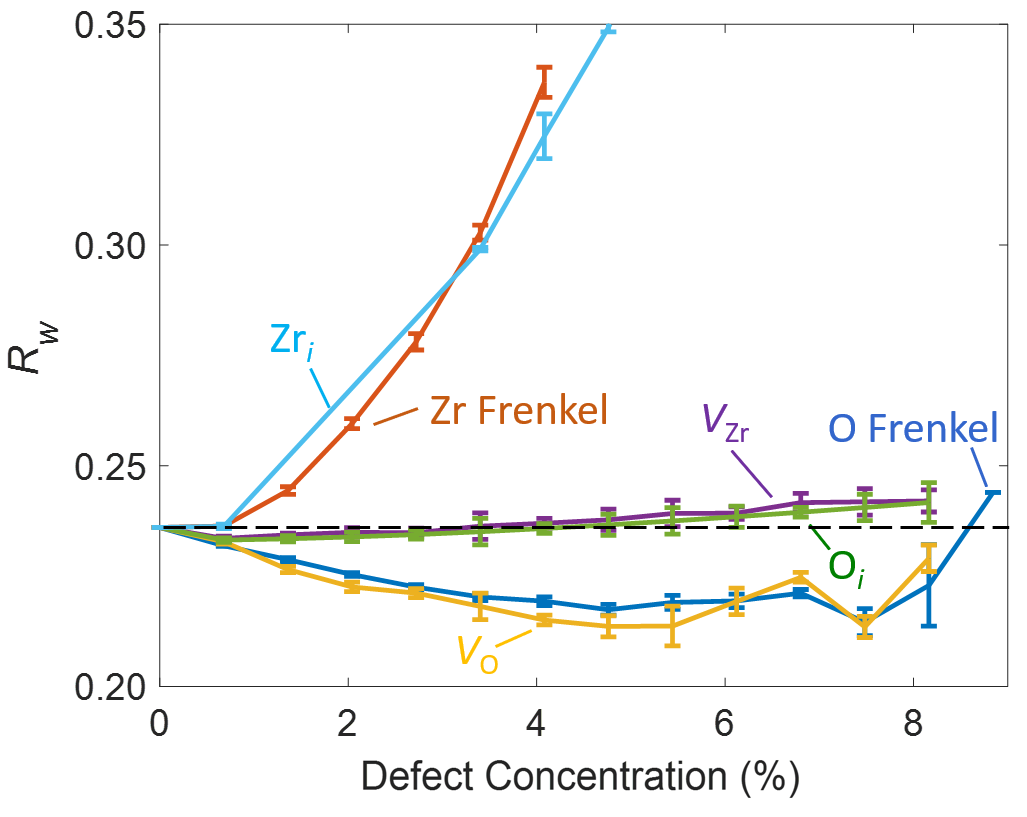}
    \caption{Variation of $R_w$ with defect concentration for tetragonal Zr$\text{O}_2$. The error bars represent the standard deviation from ten randomly-generated structures for each defect type and concentration. }\label{fig:ZrO2pdf}
\end{figure}

The variation of $R_w$ for different point defect concentrations in tetragonal Zr$\text{O}_2$ is shown in Figure~\ref{fig:ZrO2pdf}. For Zr interstitials and Zr Frenkel pairs, $R_w$  increase monotonically as the concentration increases, while $R_w$ for Zr vacancies and O interstitials remains almost unchanged from the perfect crystal value of 0.236 (plotted as a dashed horizontal line). For O vacancies and O Frenkel pairs, $R_w$ reaches a minimum at a concentration of 4.6\%. These results suggests that O point defects are the dominant defect types in MWR-grown Zr$\text{O}_2$ and that Zr-related defects are unlikely to exist. 

\begin{figure}[t]
    \centering
    \includegraphics[scale=0.31]{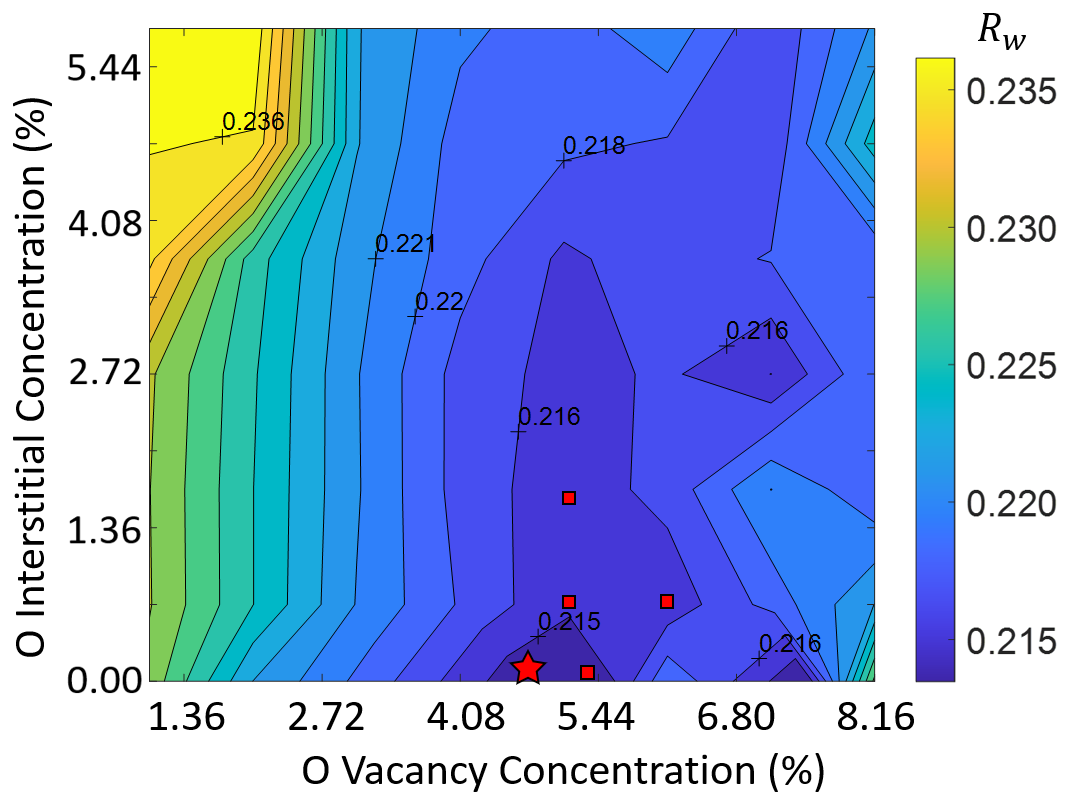}
    \caption{\label{fig:contour_zr}Variation of $R_w$ with O defect concentration for tetragonal Zr$\text{O}_2$. The red markers indicate the five lowest $R_w$ values. The lowest $R_w$, the red star, results from the PDF refinement of a structure model with 4.8\% O vacancies.}
\end{figure}

Because only O defects contribute to lowering $R_w$ in the PDF refinements for the structures in Fig.~\ref{fig:ZrO2pdf}, we built structures with different combinations of O vacancies and O interstitials and refined them against the experimental PDF. The resulting two-dimensional $R_w$ contour is shown in Fig.~\ref{fig:contour_zr}. The lowest $R_w$ (0.214) results for the structure with 4.8\% O vacancies and no O interstitials (indicated by the red star marker in Fig.~\ref{fig:contour_zr}). The refinement parameters associated with this structure are listed in Table S6 of the Supplemental Material~\cite{SM}. Four other defected tetragonal Zr$\text{O}_2$ structures give a similar $R_w$ (within 1\% of 0.214) after the PDF refinement (indicated by the red square markers in Fig.~\ref{fig:contour_zr}). All of these structures have the O vacancy as the dominant defect type, followed by the O interstitial. It should be noted that some structures leading to low $R_w$ values are not included because the refined ADPs are not physical~\cite{adp}.

\begin{figure*}[h]
    \centering
    \includegraphics[scale=0.47]{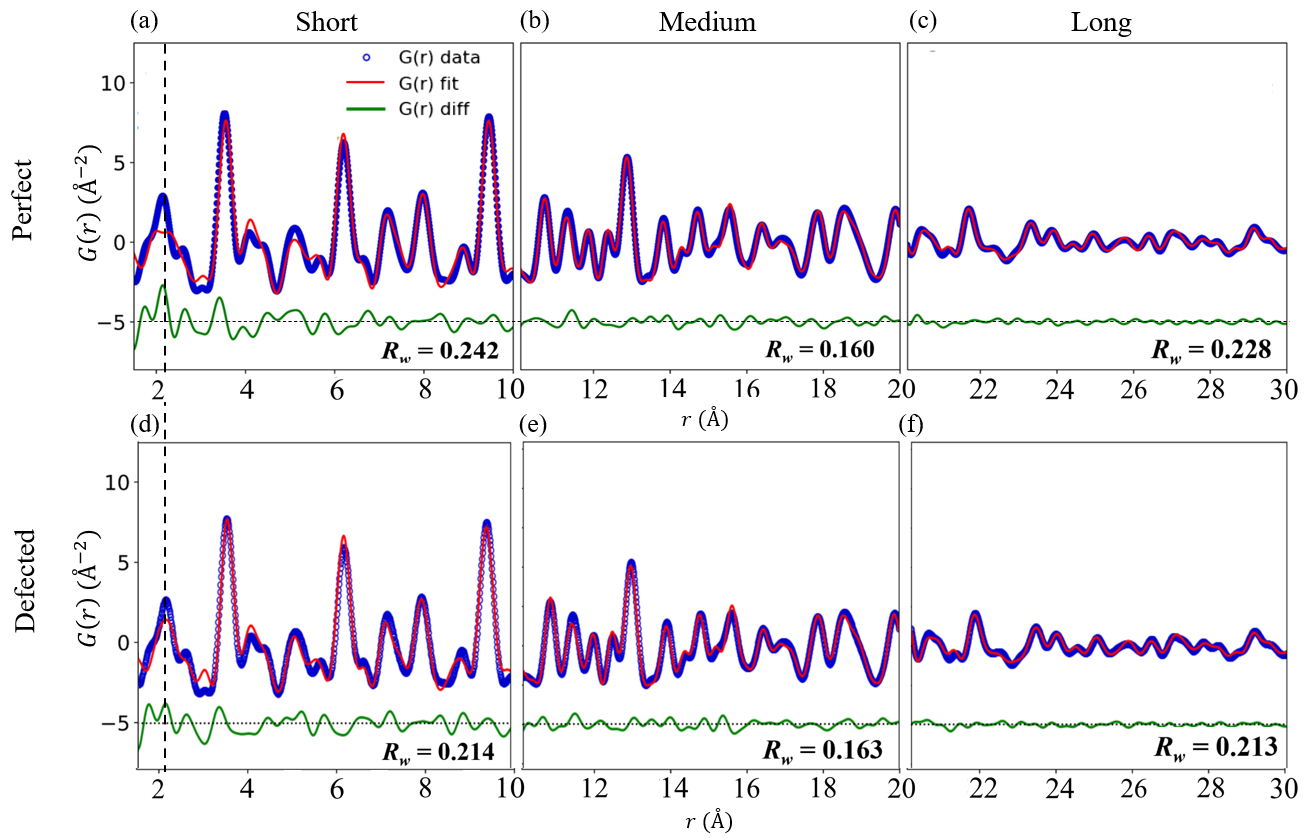}
    \caption{\label{fig:splitted_zr}PDF refinement for tetragonal Zr$\text{O}_2$ separated into short, medium, and long range order. The green curves are the differences between the refined and experimental PDFs. (a)-(c) are the refined PDFs using the perfect structure and (d)-(f) are the refined PDFs using the defected structure with the lowest $R_w$.}
\end{figure*}

The PDFs of the perfect tetragonal phase and the defected structure with the lowest $R_w$ (both refined against the same experimental PDF) are split into short, medium, and long range order in Figs.~\ref{fig:splitted_zr}(a)-\ref{fig:splitted_zr}(f). The refinement to the MWR-grown tetragonal phase using the defected structure model shows improvement in the first peak at 2.1 \AA~(vertical dashed line), which corresponds to the nearest-neighbor Zr$\textendash$O distance [see Fig. S3(b) of the Supplemental Material~\cite{SM}]. The short range $R_w$ decreases from 0.242 to 0.214, suggesting that the defected model with mainly oxygen vacancies describes the local atomic structure. The medium range $R_w$ slightly increases from 0.160 to 0.163 and the long range $R_w$ decreases from 0.228 to 0.213. This observation indicates that despite the local disorder, the atomic structure of the MWR-grown Zr$\text{O}_2$ remains tetragonal. The introduction of O vacancies and interstitials makes the local atomic environment more similar to that in the MWR-grown Zr$\text{O}_2$. 

From Table~\ref{tab:TiO2&ZrO2}, Zr interstitial has the lowest formation energy based on the SMTB-Q potential, followed by O vacancy, O interstitial, and Zr vacancy. The low ends of the DFT ranges indicate an order of O vacancy,  O interstitial, Zr interstitial, and Zr vacancy. Similar to the discussion about TiO$_2$, differences compared to the result of the PDF refinement may be a result of the finite experimental temperature and the presence of MWR. 

Our conclusion that O defects are present in MWR-grown Zr$\text{O}_2$ is consistent with a recent study by Nakamura et al$.$ on defect-mediated phase transitions in Zr$\text{O}_2$ under MWR~\cite{nakamura2021linking}, where the authors used the same experimental PDF as this work. In addition to their experimental work, they generated 96 atom perfect structures and then created one or two O vacancies or interstitials. Each structure was relaxed using DFT and then refined against the experimental PDF of their MWR-grown Zr$\text{O}_2$. They found that the tetragonal phase with 2.1\% O vacancies best approximated the MWR-grown Zr$\text{O}_2$, which is lower than the concentration we obtained (5.1\%). While DFT provides more accurate structures and energies than the SMTB-Q potential, the accessible system sizes and defect configurations are limited by the high computational cost. Furthermore, the periodicity of the defects in their small DFT computational cell, which has a minimum side length of 7 \AA, will have a more pronounced effect than in our simulation box, which contains 1470 atoms and a minimum box side length of 26 \AA (Table~\ref{tab:setup}). 

%\clearpage
\subsection{\label{Uiso}Lattice constants and atomic displacement parameters from PDF analysis}

The refined PDFs provide additional information beyond the atomic structure. We now examine the lattice constants and ADPs %atomic displacement parameters (ADP, $U_{iso}$)
of the perfect and defected structures extracted from the refined PDFs and compare the values with other experimental and modeling results. We do not refine the anisotropic ADPs because the crystallographic resolution $Q_{max}$ of the experiments we are analyzing is not high enough (16 \AA$^{-1}$). Previous PDF analysis of TiO$_2$ refined the anisotropic displacement parameters when the $Q_{max}$ was larger than 22 \AA$^{-1}$ \cite{dambournet2011combining,sanchez2019phase,hua2015morphology} or when the radiation source was neutrons~\cite{playford2020variations}. Our attempt to refine anisotropic ADPs did not yield physical ADP values or meaningful improvement of $R_w$ for both TiO$_2$ and ZrO$_2$ (See Tables S7 and S8 of the Supplemental Material). Values reported for the perfect TiO$_2$ structure are from the PDF measurement of an anatase powder sample using perfect anatase as the structure model~\cite{nakamura2020experimental}. There is no equivalent result for the perfect tetragonal ZrO$_2$ structure. Values for the defected TiO$_2$ and ZrO$_2$ structures are averaged over the low $R_w$ structures identified by red markers in Figs.~\ref{fig:contour}(a)-\ref{fig:contour}(e) and~\ref{fig:contour_zr}.

The anatase Ti$\text{O}_2$ and tetragonal Zr$\text{O}_2$ lattice constants obtained directly from energy minimization with the SMTB-Q potential are compared with the PDF refinement of the lowest $R_w$ structures in Table~\ref{tab:in-situ_lattice}. In Ti$\text{O}_2$, the lattice constant $a$ in the perfect structure expands from 3.809 \AA~to 3.905 \AA~after introducing defects, while the lattice constant $c$ contracts from 9.597 \AA~to 9.364~\AA. From the PDF refinement, $a$ also expands from 3.772 \AA~to 3.793 \AA~after introducing defects, while $c$ changes by only 0.001 \AA. In Zr$\text{O}_2$, both the lattice constants $a$ and $c$ contract after inserting defects into the structures ($a$ from 3.681~\AA~to 3.646~\AA~and $c$ from 5.285~\AA~to 5.205~\AA).

The ADP, which has the dimension of length squared, quantifies static atomic displacements away from the perfect structure and thermal vibrations about the equilibrium positions~\cite{dunitz1988interpretation}. We will present ADPs obtained from the PDF refinements and atomic mean squared displacements (MSDs) calculated from MD simulations at 300 K. While both of these quantities are impacted by local disorder and atomic vibrations, there are five factors that challenge a quantitative comparison.

(i) Thermal vibration versus static disorder. The ADPs obtained from crystal diffraction experiments provide time- and lattice-averaged probability density functions that contain contributions from thermal vibrations and static disorder, which cannot be distinguished at a given temperature~\cite{dunitz1988interpretation}. The MSD, on the other hand, only accounts for thermal vibrations around the equilibrium positions.

(ii) The distribution of MSD amplitude across atoms. When a percentage of atoms of the same type have a larger MSD than the rest, fitting to one ADP can overestimate the average atomic displacement~\cite{egami2003underneath}.

(iii) Anisotropic atomic displacements. The ADPs are refined isotropically by assuming that the vibrational amplitudes are equal in all directions~\cite{carugo2018atomic}. This assumption is made to avoid refining more than one variable per atom type due to the scarcity of diffraction data. 

(iv) Correlated atomic displacements. When two atoms are bonded, they have a tendency to move in a correlated fashion. Their displacement relative to each other is overestimated by the ADP if they move in the same direction, and is underestimated if the motions are anti-correlated~\cite{egami2003underneath,jeong1999measuring,jeong2003lattice}. 

(v) Crystallographic resolution. The accuracy of the ADP can be affected by the experimental resolution. If the resolution is insufficient, it is not possible to identify alternative atomic positions, which will be compensated by an increase of the ADP~\cite{carugo2018atomic}.  

\begin{figure*}[h]
    \centering
    \includegraphics[scale=0.5]{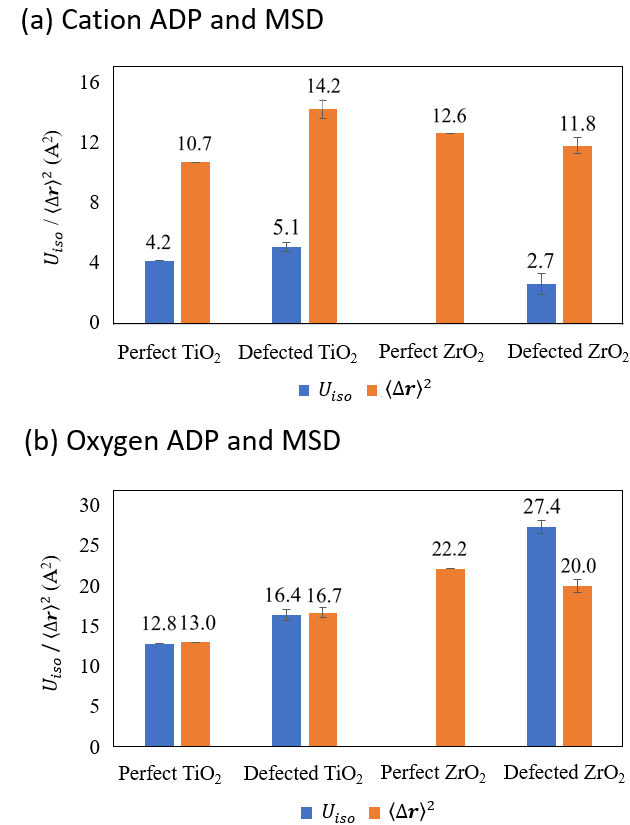}
    \caption{\label{fig:tabvi}(a) Cation and (b) O ion MSD and ADP values. The data are also provided in Table~\ref{tab:msd_tio2}}
\end{figure*}

The anatase Ti$\text{O}_2$ ADPs from the PDF refinement are provided in Table~\ref{tab:msd_tio2} and Figs.~\ref{fig:tabvi}(a) and \ref{fig:tabvi}(b). The ADPs for defected Ti$\text{O}_2$ are larger than those of the perfect structure for both Ti (by 21\%) and O (by 28\%), suggesting that the defects induce a static displacement and/or larger thermal vibrations of Ti and O atoms. As to the latter, we used MD simulations to verify that increasing the defect concentration increases the thermal vibrations. The results are shown in Figs. S4 and S5 of the Supplemental Material~\cite{SM}.

We calculated species-dependent MSDs, $\langle \Delta \bm r \rangle^2_j$ ($j=$ Ti, O),  using MD simulations at a temperature of 300~K from
\begin{equation}\label{Eq_MSD}
 \langle \Delta \bm r \rangle^2_j= \frac{1}{N_j} \left\langle \sum_{i \in N_j} (\bm r_i^{(t)} - \bm r_{i,eq})^2 \right\rangle_t.
\end{equation}
Here, $N_j$ is the number of atoms of type $j$, the summation is over all atoms of type $j$, $\bm r_i^{(t)}$ is the atomic position at time $t$, $\bm r_{i,eq}$ is the averaged atomic position, and the angle brackets denote a time average. The MD simulations were run for 10 ps (50,000 time steps). The average atomic positions and MSDs were taken over the last 15,000 time steps. We note that MSDs calculated from MD simulations will be underestimated below the Debye temperature due to the exclusion of quantum statistics~\cite{nemkevich2010molecular}. The Debye temperatures of anatase Ti$\text{O}_2$ and tetragonal Zr$\text{O}_2$ are 520 K~\cite{howard1991structural} and 515 K~\cite{lawless1983thermal}. 

\begin{table}[h]
\caption{\label{tab:in-situ_lattice} Lattice constants of perfect and defected anatase Ti$\text{O}_2$ and tetragonal Zr$\text{O}_2$ obtained directly from energy minimization and from PDF refinement. The perfect phase values are refined using perfect structure models and the defected phases are refined using the low $R_w$ structures represented by red markers in Figs.~\ref{fig:contour}(a)-\ref{fig:contour}(e) and~\ref{fig:contour_zr}. The numbers in the brackets are the standard deviations among those structures. An experimental PDF of perfect tetragonal Zr$\text{O}_2$ is not available.}
\vspace{2 mm}
\begin{ruledtabular}
\begin{tabular}{ccccc}
\multirow{2}{*}{} & \multicolumn{2}{c}{\textbf{Minimization}} & \multicolumn{2}{c}{\textbf{PDF refinement}} \\ \cline{2-5} 
 & Perfect & Defected  & Perfect & Defected          \\ \hline
Ti$\text{O}_2$ $a$ (\AA)  &3.809   &3.905(0.008)    &3.772 & 3.793(0.001)\\
Ti$\text{O}_2$ $c$ (\AA) &9.597    &9.364(0.016)     &9.491 & 9.492(0.001)\\ 
Zr$\text{O}_2$ $a$ (\AA)  &3.681    &3.646(0.010)    & -  &3.554(0.001)\\ 
Zr$\text{O}_2$ $c$ (\AA)  &5.285    &5.205(0.018)    & -  &5.097(0.001)\\ 
\end{tabular}
\end{ruledtabular}
\end{table}

The results for the perfect and defected TiO$_2$ structures are provided in Table~\ref{tab:msd_tio2}. The Ti and O MSDs in the perfect structure are 0.0107 \AA$^2$ and 0.0130 \AA$^2$, with an O/Ti ratio of 1.2. The MSDs of Ti and O in the defected structure are larger, 0.0142 \AA$^2$ and 0.0167 \AA$^2$, with no significant change of the O/Ti MSD ratio. This result is consistent with the ADPs and confirms that the defects induce a larger thermal vibration. For both the perfect and defected structures, the O MSDs are comparable to the O ADP values, while the Ti MSDs are more than twice as large as the Ti ADPs, leading to larger O/Ti ratios based on the ADP. 

Jongmanns et al.~\cite{jongmanns2020element} directly compared experimental O/Ti ADP ratio from \emph{in-situ} PDF measurement during flash sintering of rutile Ti$\text{O}_2$~\cite{yoon2018measurement} to MSD ratios calculated by MD simulations with Frenkel defects. The experimentally measured ADP O/Ti ratio is 1.51 $\textendash$ 2.02~\cite{yoon2018measurement}, which, similar to our results, is larger than their calculated MSD O/Ti ratio of 0.99 $\textendash$ 1.23~\cite{jongmanns2020element}.

The ADPs for tetragonal Zr$\text{O}_2$ are compared to the MSDs in Table~\ref{tab:msd_tio2}. For the defected structures, the MSD of Zr (0.0118 \AA$^2$) is higher than the ADP (0.0027 \AA$^2$), similar to the anatase Ti$\text{O}_2$ results. The O MSDs, however, are larger than the ADPs in the defected structures. In addition to the difference between how the atomic thermal displacements are accounted for in the MSD and ADP, which is discussed in more detail below, the larger discrepancies in tetragonal Zr$\text{O}_2$ may also be related to the performance of the SMTB-Q potential for Zr$\text{O}_2$ compared to Ti$\text{O}_2$ (Table~\ref{tab:elastic}). The MSDs in the defected structure are lower than those in the perfect structure, contrary to Ti$\text{O}_2$. This observation could be related to the fact that pure tetragonal Zr$\text{O}_2$ is not stable at room temperature without dopants such as yttrium~\cite{shukla2005mechanisms}.

\begin{table}[h]
\caption{\label{tab:msd_tio2}A comparison of calculated MSDs using MD simulation and the ADP values ($U_{iso}$) obtained from PDF refinement for anatase Ti$\text{O}_2$ and tetragonal Zr$\text{O}_2$. The ADPs for the perfect structures are refined using the perfect structure models. The ADPs for the defected structures are refined using the low $R_w$ structures represented by red markers in Figs.~\ref{fig:contour}(a)-\ref{fig:contour}(e) and~\ref{fig:contour_zr}. The numbers in the brackets are the standard deviations among those structures. The dimensionless O/cation ratio is also reported. The experimental PDF of perfect tetragonal Zr$\text{O}_2$ is not available.}
\begin{ruledtabular}
\begin{tabular}{lcccccc}
  &\multicolumn{3}{c}{\bf $U_{iso}$ ($\times 10^{-3}$ \AA$^2$) }&\multicolumn{3}{c}{\bf ${\langle \Delta \bm r \rangle^2}$  ($\times 10^{-3}$ \AA$^2$)}\\\cline{2-7}
 &Cation &O &O/Cation &Cation &O &O/Cation \\ \hline
 $\text{Perfect Ti$\text{O}_2$}$ &4.2 &12.8 &3.0 &10.7 &13.0 &1.2 \\
 $\text{Defected Ti$\text{O}_2$}$ &5.1(0.3) &16.4(0.7) &3.2 &14.2(0.6) &16.7(0.6) &1.2\\
 $\text{Perfect Zr$\text{O}_2$}$ &- &- &- &12.6 &22.2 &1.8\\
 $\text{Defected Zr$\text{O}_2$}$ &2.7(0.7) &27.4(0.8) &10.1  &11.8(0.5) &20.0(0.8) &1.7\\
\end{tabular}
\end{ruledtabular}
\end{table}

The distinct roles played by thermal vibrations and static disorder [(i) above] is a potential source of the differences between the ADPs and MSDs. But, this effect cannot explain why the ADP (accounting for both thermal vibration and static disorder) is much lower than the MSD (accounting for thermal vibration only) for the cations. Below, we examine the distribution of MSD amplitude across atoms [(ii)] and their anisotropy [(iii)] as possible origins of the differences.

We calculated the MSD distributions for anatase TiO$_2$ and tetragonal ZrO$_2$ for the perfect structures and the defected structures with the lowest $R_w$ values. The average MSDs and their three orthogonal components, along with the standard deviation and the skewness $g_1$, which measures the asymmetry of the distribution, are provided in Table~\ref{tab:msd_dist}. 

For anatase Ti$\text{O}_2$, the standard deviation and skewness for both Ti and O in the defected structure are larger than those in the perfect structure. %., indicating that in the perfect phase, the MSD per atom has a narrower and more symmetric distribution about the average MSD. 
Introducing defects leads to a wider distribution of atomic displacements. This observation suggests a source of uncertainty in the fitted ADP compared to the true value. It does not, however, explain the differences between the ADPs and MSDs, as the Ti ADP (0.0042 \AA$^{2}$) is much lower than the Ti MSD in the perfect structure (0.011 \AA$^{2}$) (Table~\ref{tab:msd_tio2}) despite having a narrower and more symmetric MSD distribution. The MSD distribution is also not a relevant factor in tetragonal ZrO$_2$. Although the skewness of the O MSD (19.9) is much higher than the Zr MSD (2.8) in the defected structures, the O MSD value is 27\% lower than O ADP while the Zr MSD is four times the Zr ADP. 

In TiO$_2$, the MSDs components increase anisotropically when defects are introduced, with the $z$-component having the largest increase of 42\% for Ti and 37\% for O. The $z$-components of the MSDs also have a larger positive skewness (20.8 for Ti and 21.8 for O), which may be caused by the enhancement of atomic vibrations around the defect sites. In ZrO$_2$, the MSDs in the defected structures are more isotropic than that in perfect structure. The anisotropic atomic displacement, however, cannot be the main factor for the difference between the MSD and ADP for Zr and Ti. The three components of the Zr MSD in the defected structures are the most isotropic values among all Ti, Zr, and O MSDs, but the Zr MSD in defected structures has the largest difference compared to the corresponding ADP from Table~\ref{tab:msd_tio2}.

\begin{table*}[h]
\caption{\label{tab:msd_dist} MSD distributions ($\times 10^{-3}$ \AA$^2$) and skewness $g_1$ calculated from MD simulations for anatase Ti$\text{O}_2$ and tetragonal Zr$\text{O}_2$. The values in the brackets are the standard deviations ($\times 10^{-3}$ \AA$^2$).}
\begin{ruledtabular}
\begin{tabular}{llcc|cc}
 \multicolumn{2}{l}{\multirow{2}{*}{}} &\multicolumn{2}{c}{\bf Anatase Ti$\text{O}_2$}&\multicolumn{2}{c}{\bf Tetragonal Zr$\text{O}_2$}\\\cline{3-6}
 \multicolumn{2}{l}{} &Defected &Perfect &Defected &Perfect\\ \hline
 Ti/Zr & ${\langle \Delta \bm r \rangle^2}$ &13.1(5.4) &10.7(1.3) &11.8(2.5) &12.6(1.8)\\
 &$g_{1}$ &17.6 &0.3  &1.3 &0.5 \\
 &${\langle \Delta \bm r \rangle^2}_x$ &4.2(1.1) &3.4(0.6)  &3.9(1.3) &3.9(1.3) \\
 &$g_{1,x}$ &4.5 &0.3  &1.9 &0.6 \\
 &${\langle \Delta \bm r \rangle^2}_y$ &4.2(0.9) &4.0(0.8)  &3.9(1.4) &4.0(1.0) \\
 &$g_{1,y}$ &3.9 &0.2  &2.9 &0.7 \\
 &${\langle \Delta \bm r \rangle^2}_z$ &4.7(5.0) &3.3(0.7)  &3.9(1.1) &4.7(1.2) \\
 &$g_{1,z}$ &20.8 &0.8  &2.4 &0.7 \\
 O &${\langle \Delta \bm r \rangle^2}$ &16.5(8.9) &13.0(1.4)  &20.0(19.9) &22.2(2.8)\\
 &$g_{1}$ &13.9 &0.3 &18.6 &0.4 \\
 &${\langle \Delta \bm r \rangle^2}_x$ &5.5(3.5) &4.4(1.6)  &6.6(8.1) &7.4(2.3) \\
 &$g_{1,x}$ &7.8 &0.4  &23.9 &0.6 \\
 &${\langle \Delta \bm r \rangle^2}_y$ &5.4(2.4) &4.8(1.7)  &6.6(7.5) &7.7(2.4) \\
 &$g_{1,y}$ &1.7 &0.4 &24.0 &0.8 \\
 &${\langle \Delta \bm r \rangle^2}_z$ &5.6(6.4) &4.1(0.7)  &6.8(12.4) &7.1(1.4)\\
 &$g_{1,z}$ &21.8 &0.6  &27.5 &0.5 \\
\end{tabular}
\end{ruledtabular}
\end{table*}

Correlated atomic motion [(iv)] is known to sharpen the widths of PDF peaks in the low $r$ region because those atoms tend to move in-phase~\cite{jeong1999measuring}. For biomolecular crystals, it was found that rigid-body, highly correlated, anharmonic motion contributes to an underestimation of the ADPs~\cite{kuzmanic2014x}. Although the effect of correlated motion is partially accounted for by the low $r$ peak sharpening coefficient to the PDF refinement (See Sec.~\ref{pdf analysis}), it is based on a Debye model analysis for an isotropic elastic solid~\cite{jeong2003lattice}, which may behave differently compared to TiO$_2$ and ZrO$_2$. %Further analysis of the effect of correlated motion on the ADP values from PDF refinement is beyond the scope of this study. 

The crystallographic resolution of the PDF measurements [(v)] used~\cite{nakamura2017unlocking,nakamura2021linking} is 16 \AA$^{-1}$, which is sufficient in most applications~\cite{egami2003underneath}. The resolution is unlikely to be a major factor here because an insufficient resolution would lead to an overestimated ADP, while in Table~\ref{tab:msd_tio2}, the cation ADPs are lower than their MSDs. 

We conclude that while it is challenging to make direct quantitative comparison between ADPs and MSDs, the ADP is qualitatively related to the MSD and they show the same trends when introducing defects. Based on the discussion of the five factors mentioned above,  the ADP is best used as a qualitative measure to examine the static displacement and the thermal vibration of the atoms relative to their lattice sites. The ADP can also be used as an indicator of the quality of the PDF refinement. For example, an abnormally high ADP could indicate that the structure model does not represent the experimental sample well, resulting in high static disorder.

\subsection{\label{Neutron}X-ray vs. neutron PDF}

We note that O is a weaker scatterer of X-rays compared to Ti and Zr. This situation makes background subtraction challenging, because small differences in the background scaling can lead to changes in the subtracted pattern, and subsequently to the PDF~\cite{banerjee2020improved}. There could thus be higher uncertainties in the PDF peak intensities, particularly for the O-O peaks. The uncertainties in some of the refinement parameters (e.g., the oxygen ADP) could also be increased. Neutron PDF could be used to complement X-ray PDF for TiO$_2$, ZrO$_2$, or other metal oxides \cite{ren2018synchrotron} because of the comparable absolute values of the atomic scattering lengths, which are listed for O, Ti, and Zr in Table S9 of the Supplemental Material~\cite{SM}. Using neutron PDF analysis in addition to X-ray PDF analysis could thus reduce the uncertainty of the refined parameters related to oxygen and enable refinement of anisotropic ADPs, which would facilitate a better description of the atomic thermal displacements.

To evaluate the potential usefulness of neutron PDF to this study, we compared simulated X-ray and neutron PDFs for perfect and defected anatase TiO$_2$ and tetragonal ZrO$_2$ structures. The results are provided in Figs.~S6 and S7 of the Supplemental Material~\cite{SM}. Fewer features are visible in the neutron PDFs. For anatase TiO$_2$, this result is partially due to the negative scattering length of Ti, which leads to cancellation effects with the positive peaks that result from the positive scattering length of O. This effect has been observed in other materials systems \cite{dmowski1988structure}. We also find that the effect of the point defects on the X-ray and neutron PDFs are comparable.\\

\section{Conclusion}
We developed a workflow to incorporate atomistic simulations into structure model generation in PDF refinement. We applied the workflow to investigate the local defect structures and concentrations within MWR-synthesized anatase Ti$\text{O}_2$ and tetragonal Zr$\text{O}_2$. The SMTB-Q interatomic potential was selected based on its good performance in calculating material properties and its ability to generate stable defected structures (Table~\ref{tab:elastic}). 
%Defect formation energies of point defects of the two materials were calculated with satisfactory accuracy compared to DFT results (Table.~\ref{tab:TiO2&ZrO2}). 
Relaxing structure models with defects before the PDF refinement provides energetically-favorable structures. By refining the energy-minimized defected structures against the experimental PDFs, we found that Ti vacancy and Ti interstitial are the dominant point defects in anatase Ti$\text{O}_2$ [Figs.~\ref{fig:contour}(a)-\ref{fig:contour}(e) and \ref{fig:splitted}(a)-\ref{fig:splitted}(f)], while O vacancy is dominant in tetragonal Zr$\text{O}_2$ [Figs.~\ref{fig:contour_zr} and~\ref{fig:splitted_zr}(a)-\ref{fig:splitted_zr}(f)]. We examined the ADPs from the PDF refinement and MSDs from MD simulations (Table.~\ref{tab:msd_tio2}), finding that comparison is hindered by limitations of the experimental analysis. 

While we focused on structures with local disorder, the proposed workflow can also benefit PDF analysis of nanoparticles, which is challenging due to their free surfaces and various sizes and shapes~\cite{christiansen2020there}. While there has been effort to automate modeling with structures mined from databases~\cite{yang2020structure} or algorithmically generated~\cite{banerjee2020cluster}, these structures are typically bulk-like and do not capture surface reconstructions. These approaches could be further refined by incorporating an energy minimization process. The surface reconstruction can be predicted using atomistic approaches such as energy minimization and/or molecular dynamics simulation, which allow the atoms near the nanoparticle surface to relax. These calculations will identify the stability of a given nanoparticle before using it as a structure model. Our workflow can also augment characterization of perovskites, where RMC is a popular approach due to their complex structures~\cite{tucker2007rmcprofile, krayzman2008simultaneous,szczecinski2014local}. By relaxing structures with energy minimization, a large simulation cell can be modeled with RMC while preserving a physical configuration.

\section*{Acknowledgements}

We thank Dr. Nathan Nakamura and Dr. Simon Billinge for helpful discussions. All authors acknowledge support from the Defense Advanced Research Projects Agency under award AIRA HR00111990030. B.R.-J. acknowledges support from the Army Research Office Young Investigator Program under contract number W911NF1710589.

%\clearpage
\appendix
\section{\label{app:prop pred}Property predictions}
\subsection{\label{sec:level2}Lattice constants}

To calculate the lattice constants, we ran MD simulations in the isothermal-isobaric ensemble for 10 ps at a temperature of 300 K and zero pressure. We fix the box angles and apply known crystallographic constraints. For orthorhombic structures (anatase and rutile Ti$\text{O}_2$, cubic and tetragonal Zr$\text{O}_2$), the angles are fixed at $90^{\circ}$. For monoclinic Zr$\text{O}_2$, the angles are fixed at $\alpha = 90^{\circ}$, $\beta = 99.1^{\circ}$, and $\gamma = 90^{\circ}$~\cite{wang1999crystal}. The box dimensions are output every ten timesteps as they fluctuate to maintain the set conditions. The average box dimensions are calculated using the output of the last 4 ps. The lattice constants are then obtained by dividing by the number of unit cells in each direction.

\subsection{\label{sec:level2}Elastic constants}

In linear elasticity, the stress components $\sigma$ may be expressed in terms of the strain components $\epsilon$,
\begin{equation}
 \sigma_i = c_{ij}\epsilon_j,
\end{equation}
where the $c_{ij}$ are the elastic constants of the homogeneous crystal, and $i$, $j$ = 1, 2,..., 6~\cite{sadd2009elasticity}. The elastic constant matrix is symmetric (i.e., $c_{ij}$ = $c_{ji}$), which reduces the number of independent coefficients from 36 to 21. The elastic constants are calculated at zero temperature. The crystal structure is first relaxed to zero pressure using a conjugate gradient algorithm by iteratively adjusting the box volume and the atomic coordinates to minimize the system energy. A compressive strain $\epsilon_{xx}$ is then applied to the simulation box, followed by a minimization process at constant volume to relax the atomic positions. The six stress components are then calculated from the deformed structure. The same procedure is repeated for the other five independent strain components ($\epsilon_{yy}$, $\epsilon_{zz}$, $\epsilon_{xy}$, $\epsilon_{yz}$, $\epsilon_{xz}$) with the same magnitude. The full stress matrix for compressive deformation $\sigma_{ij}^{comp}$ is thus obtained. The same steps are then taken under tensile deformation to calculate $\sigma_{ij}^{ten}$. The elastic constant matrix elements are obtained from
\begin{equation}
 c_{ij} = \frac{1}{2}\left(\frac{\sigma_{ij}^{comp}}{\epsilon_{ij}}+\frac{\sigma_{ij}^{ten}}{\epsilon_{ij}}\right).
\end{equation}
The bulk modulus $B$ and shear modulus $G$ are calculated using the Voigt approximation~\cite{toher2017combining}:
\begin{equation}
B_{\mathrm V} = \frac{1}{9}[(c_{11}+c_{22}+c_{33})+2(c_{12}+c_{23}+c_{13})], \label{E-BV}
\end{equation}
\begin{eqnarray}
G_{\mathrm V} = &&\frac{1}{15}[(c_{11}+c_{22}+c_{33})-(c_{12}+c_{23}+c_{13})]\nonumber\\
&&+\frac{1}{5}[(c_{44}+c_{55}+c_{66})].\label{E-GV}
\end{eqnarray}

\subsection{\label{sec:level2}Defect formation energy}
 The defect formation energies are calculated at zero temperature using energy minimization. A perfect crystal is first relaxed to zero pressure and its potential energy, $E_{perfect}$, is recorded. Next, one type of defect (i.e., a cation or anion vacancy, interstitial, or Frenkel pair) is created by inserting and/or deleting an atom. The defected structure is then relaxed to zero pressure allowing changes to the box size and its potential energy is recorded as $E_{defected}$. The defect formation energy $E_{formation}$ is then calculated from~\cite{avdeeva2015kinetics}
\begin{equation}\label{Eqn:DFE}
E_{formation} = E_{defected} - \frac{N_{defected}}{N_{perfect}}E_{perfect},
\end{equation}
where $N_{defected}$ and $N_{perfect}$ are the number of atoms in the defected and perfect systems. For structures with a vacancy, interstial, or Frenkel pair, we have $N_{defected} = N_{perfect} - 1$, $N_{defected} = N_{perfect} + 1$, or $N_{defected} = N_{perfect}$. 

When creating a vacancy, an oxygen or cation atom is randomly deleted along with its charge. When creating an interstitial, an atom is inserted with its charge set to the average charge value of the same atom type in a perfect system. For both vacancies and interstitial, the system is thus no longer charge neutral. When creating a Frenkel pair, as the interstitial-vacancy pair is created by moving an atom to a random interstitial site along with its charge, the system remains charge neutral.

\nocite{*}

% \bibliography{reference}% Produces the bibliography via BibTeX.

\begin{thebibliography}{73}
\bibitem [{xyz()}]{xyzfiles}%
  \BibitemOpen
  \href@noop {} {}\bibinfo {note} {The .xyz file generated from the atomistic
  calculations is converted into a .cif file so that DiffPy-cmi can treat it as
  a periodic structure.}\BibitemShut {Stop}%
\bibitem [{adp()}]{adp}%
  \BibitemOpen
  \href@noop {} {}\bibinfo {note} {Zr ADPs smaller than 0.001 Å$^2$ and O ADPs
  larger than 0.03 Å$^2$ are excluded.}\BibitemShut {Stop}%
\end{thebibliography}


\end{document}

%\preprint{APS/123-QED}

\title{Supplementary Material: Pair distribution function analysis driven by atomistic simulations: Application to microwave radiation synthesized Ti$\text{O}_2$ and Zr$\text{O}_2$}% Force line breaks with \\

\author{Shuyan Zhang$^1$}
\author{Jie Gong$^1$}%
\author{Daniel Xiao$^2$}
\author{B. Reeja Jayan$^1$}
\author{Alan J. H. McGaughey$^1$}
\affiliation{%
 $^1$Department of Mechanical Engineering, Carnegie Mellon University, \\Pittsburgh, Pennsylvania 15213, USA  
}%
\affiliation{%
 $^2$Department of Materials Science and Engineering, Carnegie Mellon University, \\Pittsburgh, Pennsylvania 15213, USA  
}%

\clearpage

\maketitle
%\tableofcontents
\clearpage

\section{\label{sec:level1}Lattice constants and elastic constants}

\subsection{\label{structure model generation}Rutile and anatase Ti$\text{O}_2$}

\begin{table*}[h]
\caption{\label{tab:table3}Ti$\text{O}_2$ lattice constants (Å), elastic constants (GPa), bulk modulus (GPa), and shear modulus (GPa) calculated using SMTB-Q, ReaxFF, and the literature values.}
\begin{ruledtabular}
\begin{tabular}{ccccc|cccc}
 & &\multicolumn{3}{c}{\bf Rutile}&\multicolumn{3}{c}{\bf Anatase}\\\cline{2-9}
 &SMTB-Q &\shortstack{ReaxFF \\(Ti$\text{O}_2$/$\text{H}_2\text{O}$)}& \shortstack{ReaxFF \\(defect)} &Exp.~\cite{rahimi2016review,yao2007ab} & SMTB-Q &\shortstack{ReaxFF\\ (Ti$\text{O}_2$/$\text{H}_2\text{O}$)} &\shortstack{ReaxFF \\(defect)} &Exp.~\cite{rahimi2016review,yao2007ab}\\ \hline
 $a$ &4.623 &4.615 &4.625 &4.593 &3.809 &3.798 &3.832 &3.784\\
 $c$ &2.978 &2.973 &2.979 &2.958 &9.597 &9.550 &9.673 &9.515\\ \hline
 $c_{11}$ &277 &226 &278 &269 &341 &401 &402 &337\\
 $c_{22}$ &277 &194 &235 &269 &341 &401 &412 &337\\
 $c_{33}$ &371 &476 &515 &480 &198 &211 &245 &192\\
 $c_{12}$ &156 &330 &171 &177 &104 &87 &101 &139\\
 $c_{13}$ &138 &277 &152 &146 &103 &115 &145 &136\\
 $c_{23}$ &138 &80 &129 &146 &103 &113 &155 &136\\
 $c_{44}$ &108 &96 &81 &124 &56 &-675 &-238 &54\\
 $c_{55}$ &108 &98 &82 &124 &56 &-58 &90 &54\\
 $c_{66}$ &123 &233 &162 &192 &63 &35 &53 &60\\
 $c_{ij}$ RMSE &41.3 &78.6 &25.4 &- &17.9 &249 &106 &-\\ \hline
 $B_v$\footnote{Eq.(A3)} &199 &252 &215 &218 &167 &182 &207 &178\\
 $G_v$\footnote{Eq.(A4)} &113 &142 &109 &112 &58 &-32 &-32 &57\\
 
\end{tabular}
\end{ruledtabular}
\end{table*}

\clearpage
\subsection{\label{pdf analysis}Monoclinic, tetragonal, and cubic Zr$\text{O}_2$}

\begin{table*}[h]
\caption{\label{tab:table3}Zr$\text{O}_2$ lattice constants (Å), elastic constants (GPa), bulk modulus (GPa), and shear modulus (GPa) calculated using SMTB-Q and the literature values.}
\begin{ruledtabular}
\begin{tabular}{ccc|cc|cc}
 &\multicolumn{2}{c}{\bf Monoclinic}&\multicolumn{2}{c}{\bf Tetragonal}&\multicolumn{2}{c}{\bf Cubic}\\\cline{2-7}
 &SMTB-Q & Exp.~\cite{wang1999crystal,nevitt1988elastic}& SMTB-Q &Exp.~\cite{igawa1993crystal,kisi1998elastic} & SMTB-Q &Exp.~\cite{igawa1993crystal,kandil1984single}\\ \hline
 $a$ (Å) &5.211 &5.146 &3.681 &3.591 &5.197 &5.108\\
 $c$ (Å) &5.380 &5.313 &5.285 &5.169 &5.197 &5.108\\\hline
 $c_{11}$ &287 &358 &294 &327 &458 &401\\
 $c_{22}$ &338 &426 &294 &327 &458 &401\\
 $c_{33}$ &275 &240 &178 &264 &458 &401\\
 $c_{12}$ &123 &144 &191 &100 &61.8 &96\\
 $c_{13}$ &94.4 &67.0 &16.5 &62 &61.8 &96\\
 $c_{23}$ &108 &127 &16.5 &62 &61.8 &96\\
 $c_{44}$ &76.5 &99.1 &41.7 &59 &55.6 &56\\
 $c_{55}$ &77.1 &78.7 &41.7 &59 &55.6 &56\\
 $c_{66}$ &91.2 &130 &181 &64 &55.6 &56\\
 $c_{15}$ &31.9 &-25.9 & & & &\\
 $c_{25}$ &-4.46 &38.3 & & & &\\
 $c_{35}$ &-28.2 &-23.3 & & & &\\
 $c_{46}$ &-6.50 &-38.8 & & & &\\
 $c_{ij}$ RMSE &154.3  & &190.5 & &115.1 &\\\hline
 $B_v$\footnote{Eq.(A3)} &173 &189 &135 &152 &194 &198\\
 $G_v$\footnote{Eq.(A4)} &87 &107 &89 &83 &113 &95\\

\end{tabular}
\end{ruledtabular}
\end{table*}

\section{Defect formation energy}

\subsection{Rutile and anatase Ti$\text{O}_2$}
\begin{table*}[h]
\caption{\label{tab:TiO2}Defect formation energy (eV) for a titanium or oxygen vacancy, interstial, and Frenkel pair. }
\begin{ruledtabular}
\begin{tabular}{ccccc}
&\multicolumn{2}{c}{\bf Rutile}&\multicolumn{2}{c}{\bf Anatase}\\\cline{2-5}
 &SMTB-Q &DFT &SMTB-Q &DFT \\ \hline
 $\text{O Vacancy}$ &1.59 &0.80$\textendash$4.41~\cite{morgan2010intrinsic} &1.85 &0.70$\textendash$4.23~\cite{morgan2010intrinsic}\\
 $\text{Ti Vacancy}$ &3.83 &-- &2.93 &-0.7$\textendash$5.80~\cite{na2006first}\\
 $\text{O Interstitial}$ &5.51 &-- &4.37 &0.60$\textendash$4.30~\cite{na2006first}\\
 $\text{Ti Interstitial}$ &0.06 &0.26$\textendash$7.50~\cite{morgan2010intrinsic} &1.07 &0.70$\textendash$7.74~\cite{morgan2010intrinsic}\\
 $\text{O Frenkel Pair}$ &7.18 &-- &5.38 &--\\
 $\text{Ti Frenkel Pair}$ &3.45 &1.98$\textendash$3.84~\cite{he2005ab} &3.86 &--\\
\end{tabular}
\end{ruledtabular}
\end{table*}

\subsection{Monoclinic, tetragonal, and cubic Zr$\text{O}_2$}
\begin{sidewaystable*}
\caption{
Defect formation energy (eV) for a zirconium or oxygen vacancy, interstitial, and Frenkel pair.
% The defect formation energy for a zirconium or oxygen vacancy, interstial, and Frenkel pair in comparison with the BM potential~\cite{dwivedi1990computer} and DFT results~\cite{zhang2011first,youssef2012intrinsic,zheng2007first,dong2017computational,liu2014investigation}. The range of DFT results contains discrete values in between, where defects at different charge states and/or oxidation conditions are considered. The unit of defect formation energy is eV.
}\label{tab:ZrO2-Vacancy}
\begin{ruledtabular}

\begin{tabular}{lccc|ccc|ccc}
                                                            & \multicolumn{3}{c|}{Cubic}                                                                                                                                                                                        & \multicolumn{3}{c|}{Tetragonal}                                                               & \multicolumn{3}{c}{Monoclinic}                                                     \\ \cline{2-10} 
                                                            & SMTB-Q                                                                                                       & \multicolumn{1}{c}{DFT}                                                                  & BM     & SMTB-Q & \multicolumn{1}{c}{DFT}                                                     & BM     & SMTB-Q & \multicolumn{1}{c}{DFT}                                               & BM \\ \hline
\multicolumn{1}{l|}{O Vacancy}                              & \multicolumn{1}{c}{4.08}                                                                                     & \multicolumn{1}{c}{-1.01$\textendash$6.15~\cite{zhang2011first}}   & 14.77  & 4.42   & -0.76$\textendash$6.10   ~\cite{youssef2012intrinsic} & 15.62  & 3.82   & -4.10$\textendash$6.26   ~\cite{zheng2007first} & -- \\
\multicolumn{1}{l|}{Zr Vacancy}      & 10.22                                                                                                        & 15.27$\textendash$16.53   ~\cite{zhang2011first}                   & 86.06  & 9.63   & 6.11   ~\cite{youssef2012intrinsic}                   & 85.04  & 8.29   & 5.71$\textendash$16.53   ~\cite{zheng2007first} & -- \\
\multicolumn{1}{l|}{O Interstitial}                         & 12.41\footnote{$\langle110\rangle$ site},   7.82\footnote{octahedral site} & -5.28$\textendash$15.70~\cite{zhang2011first,liu2014investigation} & -9.34  & 7.74   & 1.79   ~\cite{youssef2012intrinsic}                   & -10.42 & 9.93   & 6.64$\textendash$8.90   ~\cite{zheng2007first}  & -- \\
\multicolumn{1}{l|}{Zr Interstitial} & 1.29                                                                                                         & -7.47$\textendash$2.10   ~\cite{zhang2011first}                    & -65.91 & 1.85   & 1.96   ~\cite{youssef2012intrinsic}                   & -67.41 & 1.41   & -7.08$\textendash$3.61~\cite{zheng2007first}    & -- \\
\multicolumn{1}{l|}{O Frenkel}                         & 29.59$\rm^a$, 11.86$\rm^b$                                                                                   & 1.87$\textendash$5.30   ~\cite{zhang2011first}                     & 5.43   & 12.12  & 4.11$\textendash$7.41   ~\cite{youssef2012intrinsic}  & 5.20   & 13.91  & 4.26   ~\cite{dong2017computational}            & -- \\
\multicolumn{1}{c|}{Zr Frenkel} & 11.40                                                                                                        & 8.86$\textendash$17.37   ~\cite{zhang2011first}                    & 20.15  & 11.39  & 8.07$\textendash$18.57   ~\cite{youssef2012intrinsic} & 17.63  & 10.99  & 8.75   ~\cite{dong2017computational}            & --

\end{tabular}
\end{ruledtabular}

\end{sidewaystable*}

\clearpage

\section{Variation of $R_w$ values}

\begin{figure}[h]
    \centering
    \includegraphics[scale=0.70]{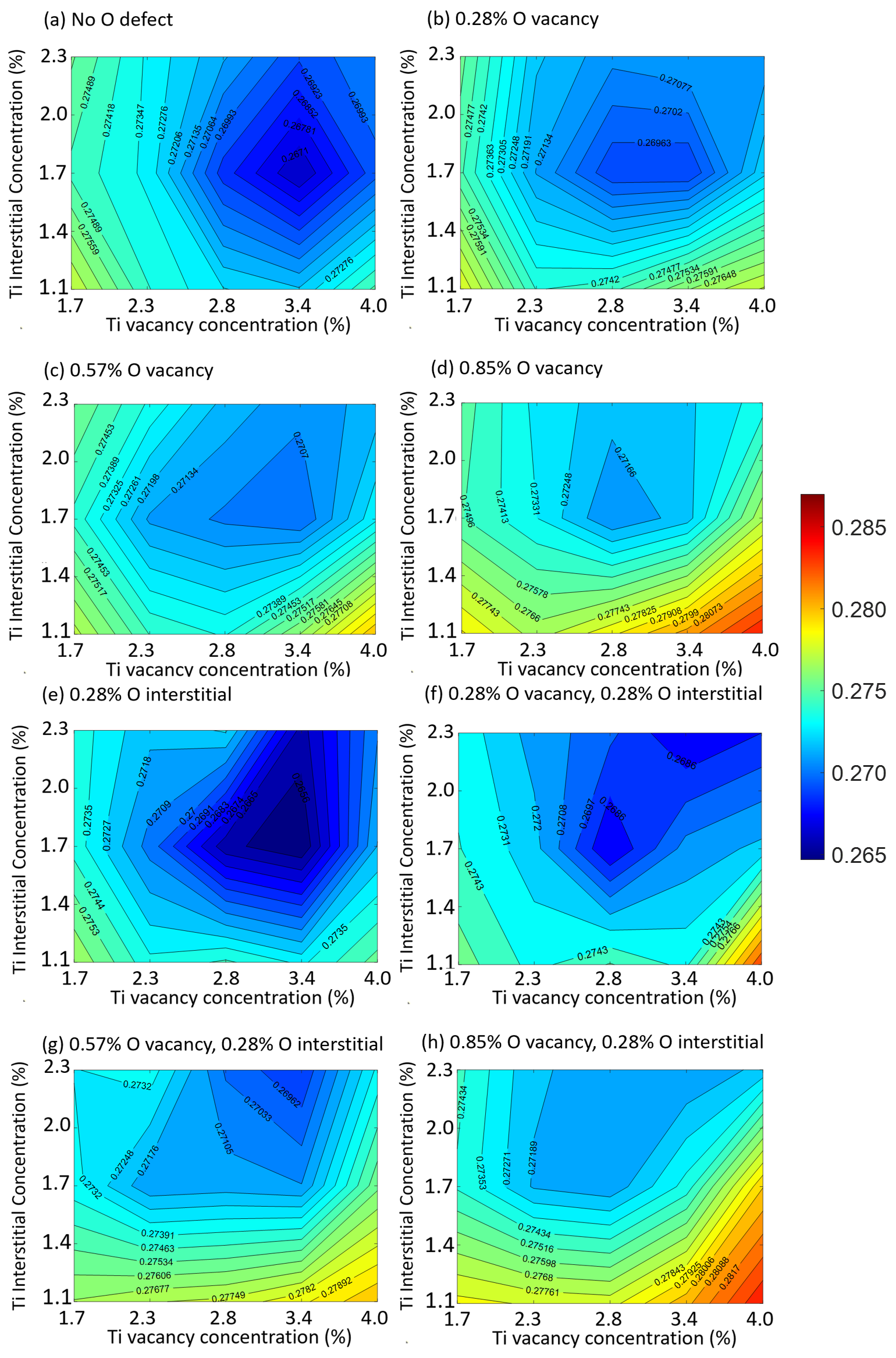}
    \caption{\label{fig:contour}Variation of $R_w$ with Ti defect concentration. The lowest $R_w$ results from the PDF refinement of a structure model with  3.4\% Ti vacancies, 1.7\% Ti interstitial and 0.28\% O interstitial.}
    \label{fig:pdf1}
\end{figure}

\begin{figure}[h]
    \centering
    \includegraphics[scale=0.7]{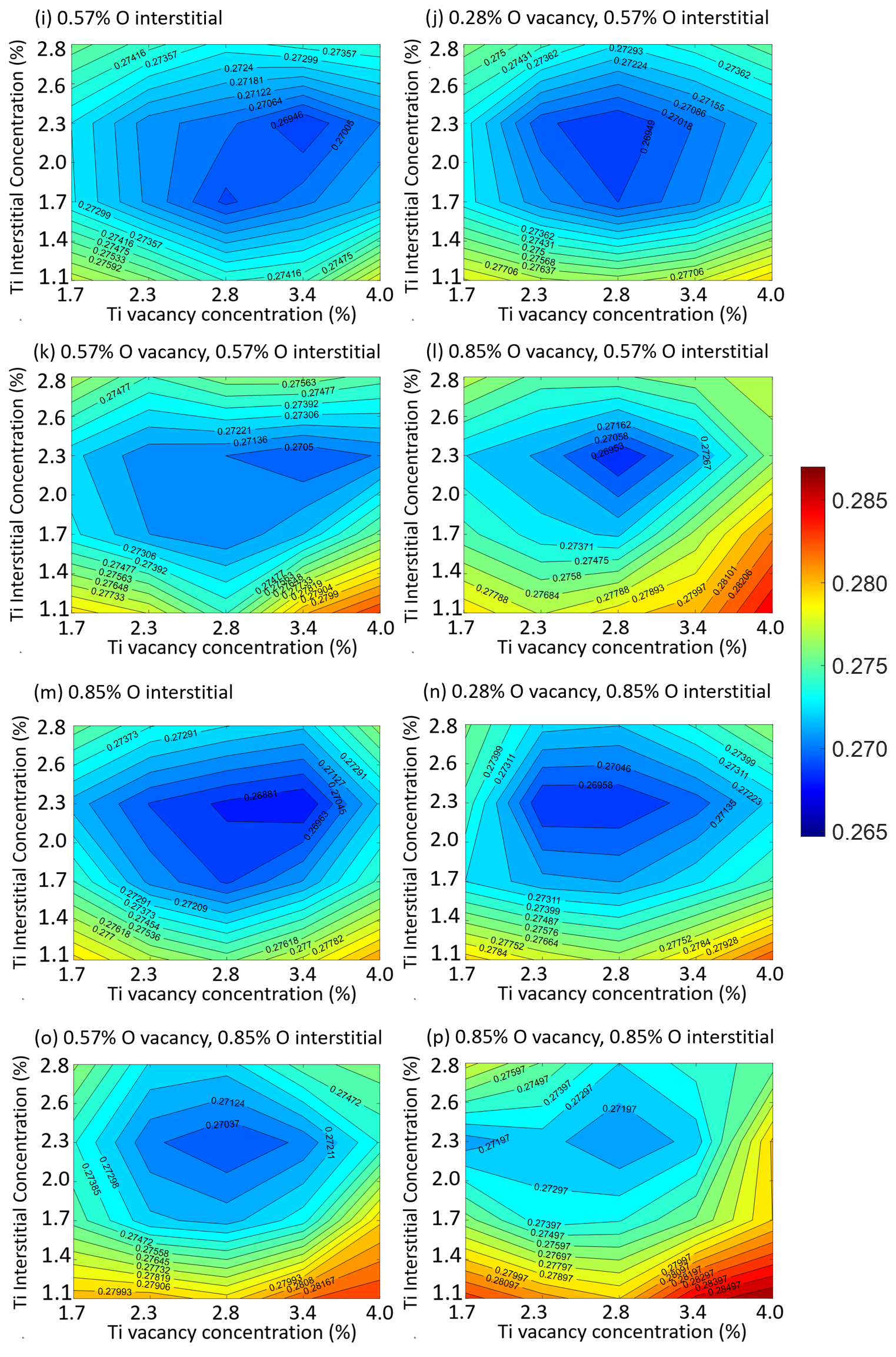}
    \caption{\label{fig:contour}Variation of $R_w$ with Ti defect concentration. The lowest $R_w$ results from the PDF refinement of a structure model with  3.4\% Ti vacancies, 1.7\% Ti interstitial and 0.28\% O interstitial.}
    \label{fig:pdf1}
\end{figure}

\clearpage

\section{PDF refinement parameters}

\begin{table*}[h]
\caption{\label{tab:table3}The refined parameters for the MWR-grown anatase Ti$\text{O}_2$ using perfect structure and defected structure.}
\begin{ruledtabular}
\begin{tabular}{lcc}
 &\bf Perfect & \bf Defected\\ \hline
 
 $a$ (Å) &3.788 &3.793\\
 $c$ (Å) &9.487 &9.492 \\
 Ti $U_{iso}$ (Å$^2$) &0.00665 &0.00510 \\
 O $U_{iso}$ (Å$^2$) &0.0238 &0.0164 \\
 $D_c$ (Å) &34.4 &33.0 \\
 $\delta_2$ &2.93 &2.54 \\
 $R_w$ &0.325 &0.264 \\

\end{tabular}
\end{ruledtabular}
\end{table*}

\begin{table*}[h]
\caption{\label{tab:table3}The refined parameters for the MWR-grown tetragonal Zr$\text{O}_2$ using perfect structure and defected structure.}
\begin{ruledtabular}
\begin{tabular}{lcc}
 &\bf Perfect & \bf Defected\\ \hline
 $a$ (Å) &3.551 &3.554 \\
 $c$ (Å) &5.097 &5.097 \\
 Zr $U_{iso}$ (Å$^2$) &0.00848 &0.00270 \\
 O $U_{iso}$ (Å$^2$) &0.0170 &0.0274 \\
 $D_c$ (Å) &40.5 &39.6 \\
 $\delta_2$ &1.87 &3.32 \\
 $R_w$ &0.236 &0.214 \\

\end{tabular}
\end{ruledtabular}
\end{table*}

\begin{table*}[h]\centering
\caption{\label{tab:tio2} Refined anisotropic ADPs, isotropic ADPs, and $R_w$ values for defected and perfect TiO$_2$. The refined $U_{22,\text{Ti}}$ is an order of magnitude smaller than that for the perfect structure, which is nonphysical. The $R_w$ values do not show significant improvement for either the defected or perfect structures.}
\begin{ruledtabular}
\vspace{2 mm}
\begin{tabular}{lcccc}
\multirow{2}{*}{} & \multicolumn{2}{c}{\textbf{Defected}} & \multicolumn{2}{c}{\textbf{Perfect}} \\ \cline{2-5} 
 & Anisotropic & Isotropic  & Anisotropic & Isotropic          \\ \hline
$U_{11,\text{Ti}}$ (Å$^2$) &0.00443   &    &0.00817 & \\
$U_{22,\text{Ti}}$ (Å$^2$) &$=U_{11,\text{Ti}}$    &0.00510    &0.00026 & 0.00420\\ 
$U_{33,\text{Ti}}$ (Å$^2$)  &0.00520    &    & 0.00358  &\\ \hline
$U_{11,\text{O}}$ (Å$^2$)  &0.0186    &    & 0.0242  &\\ 
$U_{22,\text{O}}$ (Å$^2$) &$=U_{11,\text{O}}$ &0.0164 &0.00732 &0.0128\\
$U_{33,\text{O}}$ (Å$^2$) &0.0141 & &0.0110 &\\
$R_w$ &0.264 &0.264 &0.101 &0.107\\
\end{tabular}
\end{ruledtabular}
\end{table*}

\begin{table}[h]\centering
\caption{\label{tab:zro2} Refined anisotropic ADPs, isotropic ADPs, and $R_w$ value of defected ZrO$_2$. The refined $U_{11,\text{O}}$ is two orders of magnitude smaller the refined $U_{33,\text{O}}$ for the defected structure, which is nonphysical. The $R_w$ value does not show a significant improvement. No experimental sample for perfect ZrO$_2$ is available. }
\begin{ruledtabular}
\vspace{2 mm}
\begin{tabular}{lcc}
\multirow{2}{*}{} & \multicolumn{2}{c}{\textbf{Defected}}  \\ \cline{2-3} 
 & Anisotropic & Isotropic         \\ \hline
$U_{11,\text{Zr}}$ (Å$^2$) &0.00442   &     \\
$U_{22,\text{Zr}}$ (Å$^2$) &$=U_{11,\text{Zr}}$    &0.00270    \\ 
$U_{33,\text{Zr}}$ (Å$^2$)  &0.00089   &    \\ \hline
$U_{11,\text{O}}$ (Å$^2$)  &0.0050    &    \\ 
$U_{22,\text{O}}$ (Å$^2$) &$=U_{11,\text{O}}$ &0.0274 \\
$U_{33,\text{O}}$ (Å$^2$) &0.1710 & \\
$R_w$ &0.203 &0.213 \\
\end{tabular}
\end{ruledtabular}
\end{table}

\clearpage

\section{Illustration of atomic separation corresponding to PDF peaks}

\begin{figure}[h]
    \centering
    \includegraphics[scale=0.60]{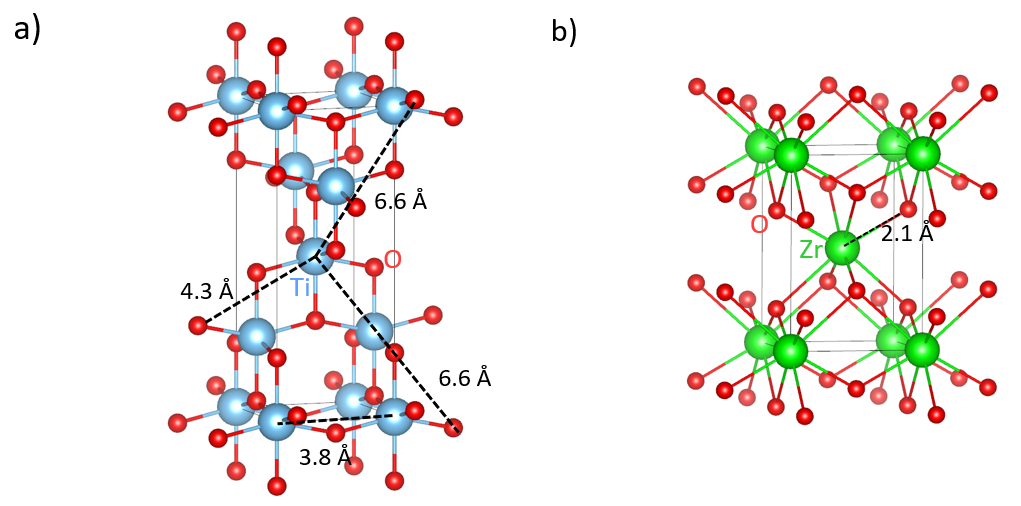}
    \caption{\label{fig:bond_dist}Atom pairs that correspond to improved peak refinement for (a) TiO$_2$ and (b) ZrO$_2$.}
\end{figure}

\clearpage

\section{The effect of defect concentration on thermal vibration}

\begin{figure*}[h]
    \centering
    \includegraphics[scale=0.6]{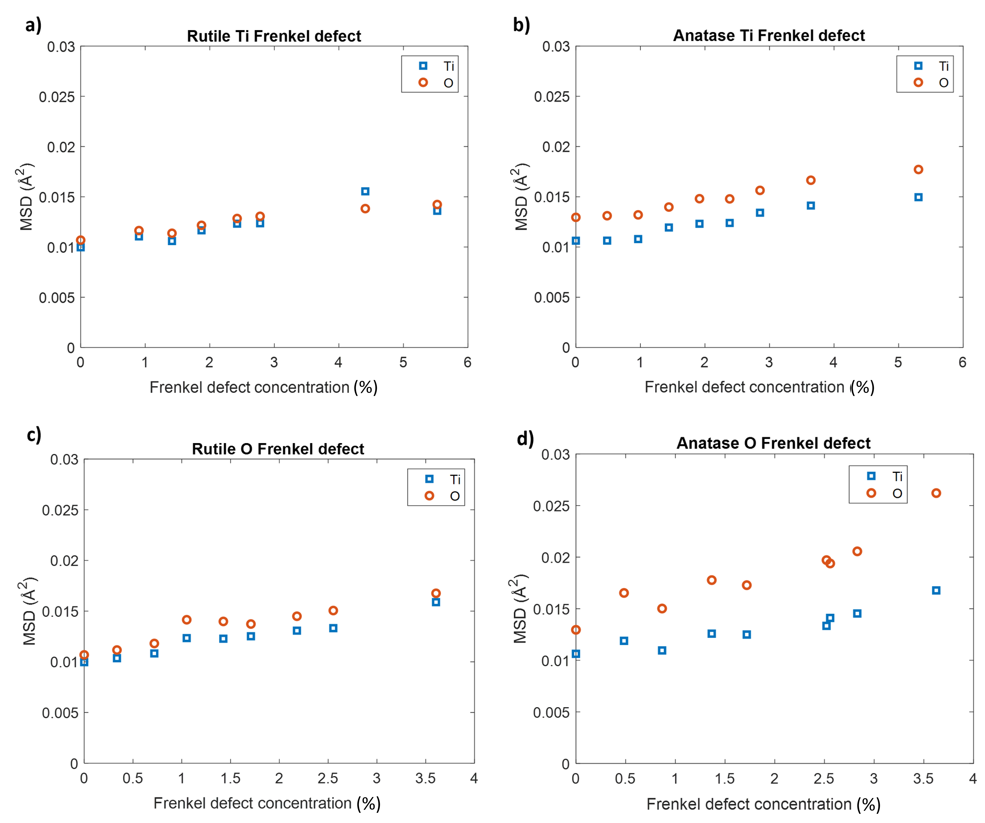}
    \caption{The element-wise MSDs of a) rutile with Ti Frenkel defects, b) anatase with Ti Frenkel defects, c) rutile with O Frenkel defects and d) anatase with O Frenkel defects.}
    \label{fig:msd1}
\end{figure*}

\begin{figure*}[h]
    \centering
    \includegraphics[width=0.9\textwidth]{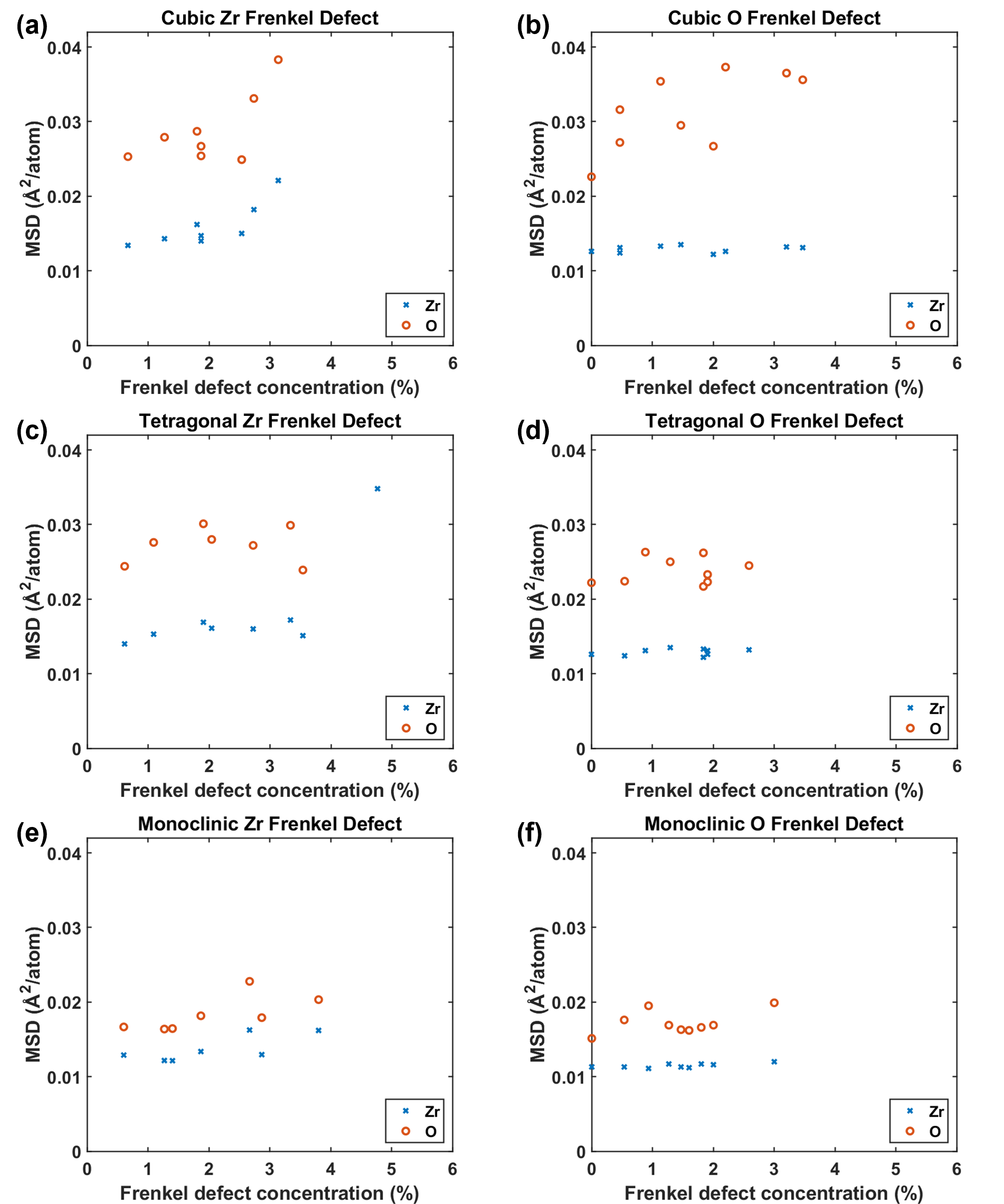}
    \caption{The element-wise MSDs of a) cubic ZrO$_2$ with Zr Frenkel defects, b) cubic Zr$\text{O}_2$ with O Frenkel defects, c) tetragonal Zr$\text{O}_2$ with Zr Frenkel defects, d) tetragonal Zr$\text{O}_2$ with O Frenkel defects, e) monoclinic Zr$\text{O}_2$ with Zr Frenkel defects and f) monoclinic Zr$\text{O}_2$ with O Frenkel defects.}
    \label{fig:msd2}
\end{figure*}

\clearpage

\section{X-ray PDF vs. neutron PDF}

\begin{table*}[h]
\caption{\label{tab:table3}The scattering factors used in DiffPy for Ti, Zr, and O.}
\begin{ruledtabular}
\begin{tabular}{lcc}
 Element &\bf X-ray ({\it e}/atom) & \bf Neutron (fm)\\ \hline
 Ti    & 22.00             & -3.37                    \\
Zr     & 39.98           & 7.16                        \\
O & 8.00              & 5.81                       \\
\end{tabular}
\end{ruledtabular}
\end{table*}

\begin{figure}[h]
    \centering
    \includegraphics[width=1\textwidth]{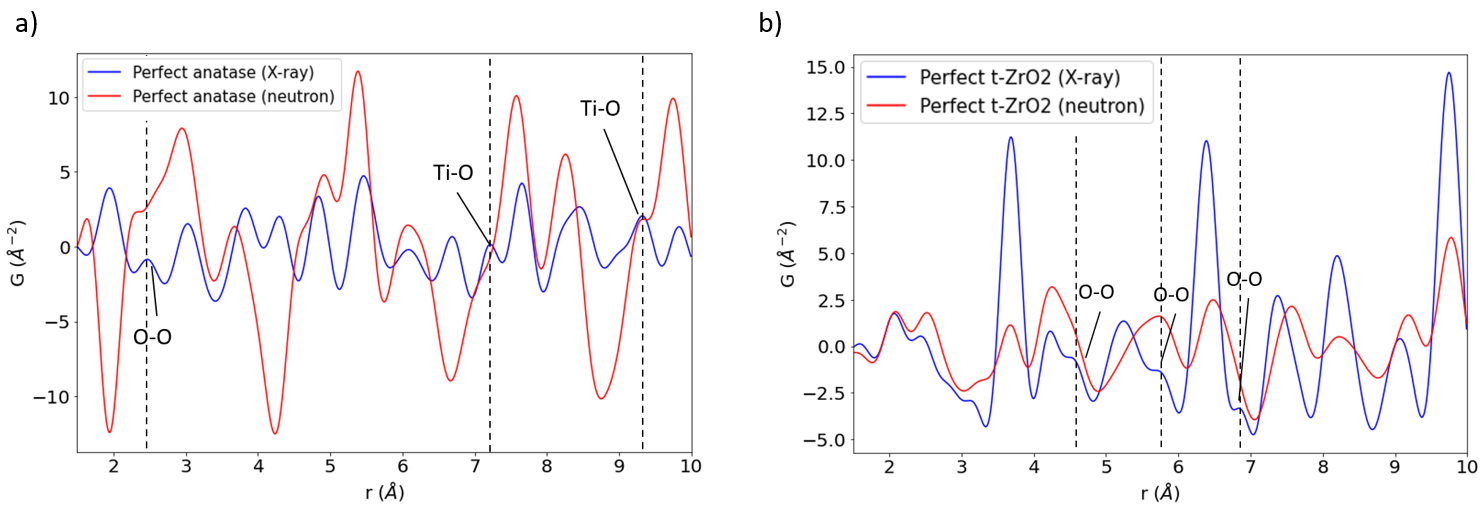}
    \caption{\label{fig:per_x_vs_n} Comparisons of X-ray (blue) and neutron (red) PDFs for perfect (a) anatase TiO$_2$ and (b) tetragonal ZrO$_2$.}
\end{figure}

The X-ray and neutron PDFs of perfect anatase TiO$_2$ and tetragonal ZrO$_2$ are shown in Fig.~\ref{fig:per_x_vs_n}. No refinement was performed and the calculation parameters were kept the same for the X-ray and neutron PDFs to enable a fair comparison. The strongly negative peaks in the neutron PDF are due to the negative neutron scattering length of Ti. These negative peaks lead to cancellation effects with the positive peaks that result from the positive scattering length of O.

\clearpage

\begin{figure}[h]
    \centering
    \includegraphics[width=1\textwidth]{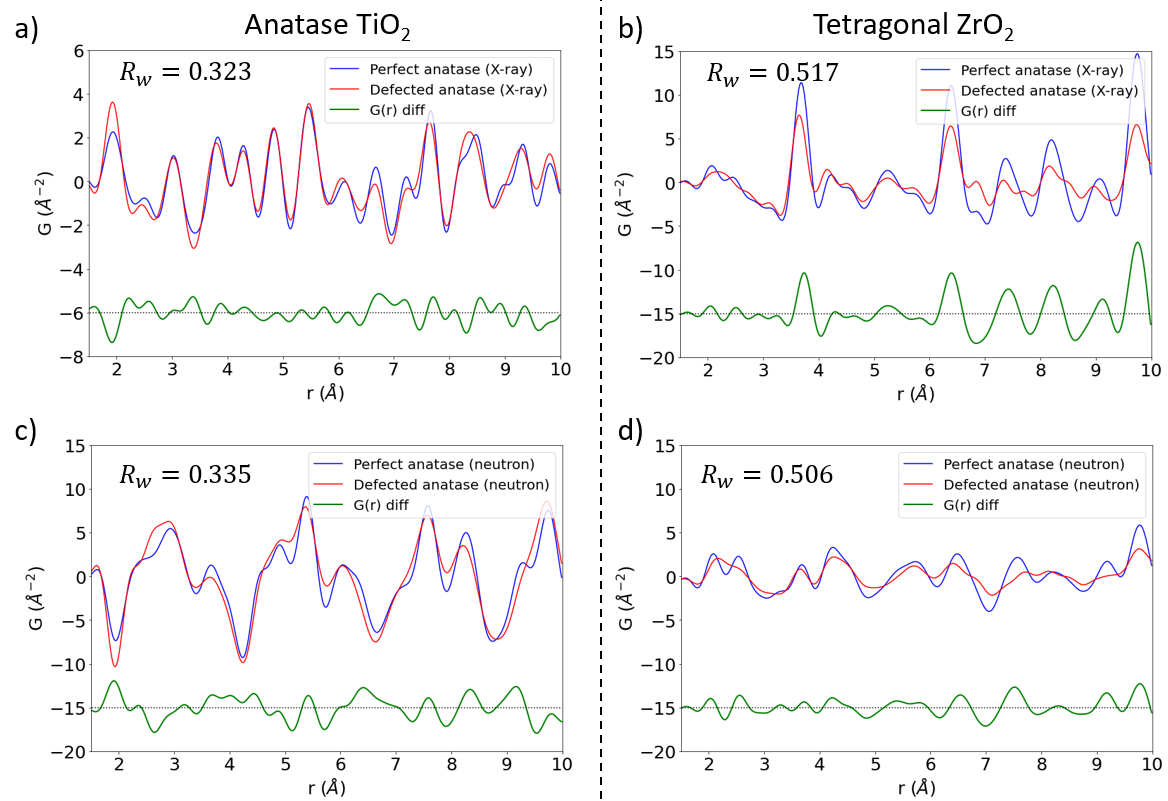}
    \caption{\label{fig:def_x_vs_n} Calculated X-ray and neutron PDFs for perfect and defected structures of anatase TiO$_2$ [(a) and (c)] and tetragonal ZrO$_2$ [(b) and (d)].}
\end{figure}

The PDFs for defected and perfect structures based on X-ray and neutron scattering are shown in Fig.~\ref{fig:def_x_vs_n}. The defected TiO$_2$ and ZrO$_2$ structures are the ones that result in the lowest $R_w$ when refined against the PDFs of the microwave radiation-grown experimental structures. The effect of the point defects on the lattice constants is removed by setting the lattice constants of the perfect and defected structures to be the same. The difference between the PDFs of the perfect and defected structures are comparable for X-rays and neutrons. Some features, such as the peaks indicated by the vertical dashed lines in Fig~\ref{fig:per_x_vs_n}, that are visible in X-ray PDFs are harder to see in the neutron PDFs.

\nocite{*}
\clearpage

\bibliography{supplement}% Produces the bibliography via BibTeX.